\documentclass[twocolumn]{aastex7}

\begin{document}

\title{The Prevalence of Bursty Star Formation in Low-Mass Galaxies at $z=1$--$7$ from H$\alpha$-to-UV Diagnostics}

\author[0009-0008-1378-4381]{Marissa N. Perry}
\affiliation{The University of Texas at Austin, Department of Astronomy 2515 Speedway, Stop C1400, Austin, TX 78712, USA}
\email{mnp944@utexas.edu}

\author[0000-0003-1282-7454]{Anthony J. Taylor}
\affiliation{The University of Texas at Austin, Department of Astronomy 2515 Speedway, Stop C1400, Austin, TX 78712, USA}
\email{anthony.taylor@austin.utexas.edu}

\author[0000-0003-2332-5505]{Óscar A. Chávez Ortiz}
\affiliation{The University of Texas at Austin, Department of Astronomy 2515 Speedway, Stop C1400, Austin, TX 78712, USA}
\email{chavezoscar009@utexas.edu}

\author[0000-0001-8519-1130]{Steven L. Finkelstein}
\affiliation{The University of Texas at Austin, Department of Astronomy 2515 Speedway, Stop C1400, Austin, TX 78712, USA}
\email{stevenf@astro.as.utexas.edu}

\author[0000-0002-9393-6507]{Gene C. K. Leung}
\affiliation{MIT Kavli Institute for Astrophysics and Space Research, 77 Massachusetts Ave., Cambridge, MA 02139, USA}
\email{gckleung@mit.edu}

\author[0000-0002-9921-9218]{Micaela B. Bagley}
\affiliation{Astrophysics Science Division, NASA Goddard Space Flight Center, 8800 Greenbelt Rd, Greenbelt, MD 20771, USA}
\affiliation{The University of Texas at Austin, Department of Astronomy 2515 Speedway, Stop C1400, Austin, TX 78712, USA}
\email{mbagley@utexas.edu}

\author[0000-0003-0531-5450]{Vital Fern\'andez}  
\affiliation{Michigan Institute for Data Science, University of Michigan, 500 Church Street, Ann Arbor, MI 48109, USA}
\email{vgf@umich.edu}

\author[0000-0002-7959-8783]{Pablo Arrabal Haro}
\affiliation{Astrophysics Science Division, NASA Goddard Space Flight Center, 8800 Greenbelt Rd, Greenbelt, MD 20771, USA}
\email{parrabalh@gmail.com}

\author[0000-0003-4922-0613]{Katherine Chworowsky}
\affiliation{The University of Texas at Austin, Department of Astronomy 2515 Speedway, Stop C1400, Austin, TX 78712, USA}
\email{k.chworowsky@utexas.edu}

\author[0000-0001-7151-009X]{Nikko J. Cleri}
\affiliation{Department of Astronomy and Astrophysics, The Pennsylvania State University, University Park, PA 16802, USA}
\affiliation{Institute for Computational and Data Sciences, The Pennsylvania State University, University Park, PA 16802, USA}
\affiliation{Institute for Gravitation and the Cosmos, The Pennsylvania State University, University Park, PA 16802, USA}
\email{cleri@psu.edu}

\author[0000-0001-5414-5131]{Mark Dickinson}
\affiliation{NSF’s National Optical-Infrared Astronomy Research Laboratory, 950 N. Cherry Ave., Tucson, AZ 85719, USA}
\email{mark.dickinson@noirlab.edu}

\author[0000-0001-7782-7071]{Richard S. Ellis}
\affiliation{University College London, Department of Physics \& Astronomy, Gower Street, London WC1E 6BT, UK}
\email{richard.ellis@ucl.ac.uk}

\author[0000-0001-9187-3605]{Jeyhan S. Kartaltepe}
\affiliation{Laboratory for Multiwavelength Astrophysics, School of Physics and Astronomy, Rochester Institute of Technology, 84 Lomb Memorial Drive, Rochester, NY 14623, USA}
\email{jeyhan@astro.rit.edu}

\author[0000-0002-6610-2048]{Anton M. Koekemoer}
\affiliation{Space Telescope Science Institute, 3700 San Martin Drive,
Baltimore, MD 21218, USA}
\email{koekemoer@stsci.edu}

\author[0000-0001-9879-7780]{Fabio Pacucci}
\affiliation{Center for Astrophysics | Harvard \& Smithsonian, 60 Garden St, Cambridge, MA 02138, USA}
\affiliation{Black Hole Initiative, Harvard University, 20 Garden St, Cambridge, MA 02138, USA}
\email{fabio.pacucci@cfa.harvard.edu}

\author[0000-0001-7503-8482]{Casey Papovich}
\affiliation{George P. and Cynthia Woods Mitchell Institute for Fundamental Physics and Astronomy, Department of Physics and Astronomy, Texas A\&M University, College Station, TX, USA}
\email{papovich@tamu.edu}

\author[0000-0003-3382-5941]{Nor Pirzkal}
\affiliation{ESA/AURA Space Telescope Science Institute}
\email{npirzkal@stsci.edu}

\author[0000-0002-8224-4505]{Sandro Tacchella}
\affiliation{Kavli Institute for Cosmology, University of Cambridge, Madingley Road, Cambridge CB3 0HA, UK}
\affiliation{Cavendish Laboratory, University of Cambridge, 19 JJ Thomson Avenue, Cambridge CB3 0HE, UK}
\email{st578@cam.ac.uk}

\begin{abstract}
We present an analysis of bursty star-formation histories (SFHs) of 346 star-forming galaxies at $1\lesssim z<7$, selected from \textit{JWST}/NIRSpec G395M and PRISM spectroscopy provided by the CEERS and RUBIES surveys. We analyze the correlation of star-formation rate vs. stellar mass (the star-forming main sequence, SFMS) for our sample and find no significant difference between the intrinsic scatter in the H$\alpha$-based SFMS and the UV-continuum-based SFMS. However, the diagnostic power of the SFMS is limited at high redshift and low stellar mass due to observational biases that exclude faint, quenched galaxies. To more directly probe star-formation variability, we examine the dust-corrected H$\alpha$-to-UV ratio, which is assumed to trace deviations a from constant SFH over the past $\sim100$~Myr. In our sample, $73^{+4}_{-4}$~\% of galaxies exhibit H$\alpha$-to-UV ratios inconsistent with a constant SFH. We do not observe any statistically significant evolution in the H$\alpha$-to-UV ratio with redshift. Additionally, lower-mass galaxies ($7\leq\text{log}(M_*/M_{\odot})<8.5$) are $30 \pm 1$\% more likely to lie above this equilibrium range---indicative of a recent ($\lesssim 100$~Myr) burst of star formation---compared to higher-mass systems ($8.5\leq\text{log}(M_*/M_{\odot})\leq10.9$). These results suggest that bursty SFHs are more common in low-mass galaxies at $z\sim 1$--$7$ and that this remains relatively stable across $\sim0.8$--$6$~Gyr after the Big Bang.
\end{abstract}


\section{Introduction} \label{sec:intro}
One of the most fundamental properties of a galaxy is the rate at which it forms stars---its star-formation rate (SFR). At any given redshift, most galaxies follow a linear relationship between log SFR versus log stellar mass ($M_{\star}$), known as the star-forming main sequence \citep[SFMS;][]{Noeske2007,Elbaz2007,Daddi2007}. Systems that lie above the intrinsic scatter of this relation are classified as ``starburst" galaxies, while those that lie below are classified as ``quiescent" galaxies. These classifications reflect an increased or decreased SFR relative to the typical value at a given stellar mass. A tight SFMS implies that galaxies of similar mass have similarly smooth and steady star-formation histories (SFHs) over cosmic time, while increased scatter suggests more stochastic or bursty SFHs \citep[e.g.,][]{Noeske2007,Hopkins2014,Shen2014,Guo2016,Asquith2018,Faisst2019,Tacchella2020,Atek2022}.

Observations increasingly support that star formation varies over cosmic time. In particular, galaxies at earlier epochs ($z\gtrsim9$) exhibit more stochastic star formation than their later-time counterparts \citep[e.g.,][]{Ciesla2024,Cole2025}. Simulations predict that this transition arises from the gradual deepening of galaxy gravitational potentials over time, which increasingly stabilize gas outflows against internal feedback mechanisms \citep[e.g.,][]{Dayal2013,Kimm2014,Tacchella2016,Wilkins2023,Hopkins2023}. With \textit{JWST} \citep{Gardner2006,Gardner2023}, such stochasticity has been proposed as one of the possible key mechanisms behind the over abundance of UV-bright galaxies observed at $z\gtrsim10$ \citep[e.g.,][]{Sun2023a,Sun2023b,Shen2023,Mason2023,Mirocha_Furlanetto2023,Kravtsov2024,Somerville2025,Carvajal-Bohorquez2025}. Indeed, the excess number density of bright galaxies observed by \textit{JWST}, and its apparent lack of evolution between $z \sim 9$ and $z \sim 14$, already challenges standard model predictions \citep[e.g.,][]{Conselice2023,Harikane2023,Finkelstein2023,Finkelstein2024,Donnan2024,Whitler2025}, motivating a closer look at the physical mechanisms behind early star formation. Thus, characterizing the burstiness of early galaxies through their SFHs is essential for understanding the extent to which stochastic star formation contributes to the observed overabundance of UV-bright galaxies.

One approach to probing burstiness is to compare time-sensitive tracers of star formation. Two widely used tracers are non-resonant hydrogen recombination lines (e.g. H$\alpha$, H$\beta$) and the UV continuum. The luminosity of Balmer lines such as H$\alpha$ (e.g., $L_{\rm{}H\alpha}$) originates from ionizing photons produced by massive, short-lived O~stars and traces star formation over $\sim$10~Myr timescales. A galaxy's UV-continuum luminosity ($L_{\rm{}UV}$) is emitted by O~stars as well as longer-lived B~stars and is traditionally assumed to trace star formation averaged over $\sim$100~Myr. However, both indicators are susceptible to several systematic effects. The H$\alpha$ luminosity depends on the retention of Lyman continuum (LyC) photons, which can vary with galaxy stellar mass, redshift, and environment due to factors such as LyC escape and dust absorption \citep[e.g.,][]{Tacchella2022-2}. Meanwhile, UV-based SFRs may be inflated by residual emission from older stellar populations, imposing a luminosity ``floor" that can obscure recent variability---particularly in galaxies with declining or bursty SFHs \citep[e.g.,][]{McClymont2025-1}. Recent modeling using Monte Carlo radiative transfer of non-ionizing photons suggests that H$\alpha$ and UV tracers are sensitive to shorter timescales than previously assumed, with H$\alpha$ tracing $\sim$7--9~Myr and the UV continuum tracing $\sim$22--31~Myr \citep{McClymont2025-1}. Moreover, both tracers require dust attenuation corrections, and the adoption of a single dust attenuation curve \citep[i.e.,][]{Calzetti2001} introduces additional uncertainty \citep[e.g.,][]{Guo2012,Broussard2019,Cleri2022,Broussard2022}.

Despite these caveats, comparing $L_{\rm{}H\alpha}$ and $L_{\rm{}UV}$ provides a powerful diagnostic of recent star-formation variability. During an episode of steady star formation, $L_{\rm{}H\alpha}$ reaches an equilibrium value much more quickly than $L_{\rm{}UV}$. Therefore, comparing both tracers allows us to detect recent variations in SFR over timescales of 10 Myr~$\lesssim~t~\lesssim$~100 Myr. Specifically, the logarithmic ratio log($L_{\rm{}H\alpha}/L_{\rm{}UV}$) (hereafter, the H$\alpha$-to-UV ratio) can be used to determine if the SFR of a given galaxy has changed significantly over the past $\sim$100~Myr \citep[e.g.,][]{Glazebrook1999,Sullivan2000,Sullivan2001,Broussard2019,Emami2019,Faisst2019,Atek2022,Broussard2022,Asada2024,Endsley2024,Clarke2024}.

The goal of this work is to use multiple SFR indicators to characterize the burstiness of star formation as a function of both redshift and stellar mass. We do this with a statistically robust sample of 346 star-forming galaxies, uniquely spanning a wide redshift range ($0.8 \leq z < 7$) and stellar mass range ($7.0 \leq \text{log}(M_*/M_{\odot}) \leq 10.9$), exceeding the scope of many prior observational studies due to the ability of {\it JWST} to probe H$\alpha$ to higher redshifts ($z\lesssim 7$) than possible from the ground. 

We organize this work as follows: In Section~\ref{sec:data} we describe \textit{JWST}/NIRSpec and \textit{JWST}/NIRCam data used in our analysis. In Section~\ref{sec:methods} we outline methods used to derive the SFR indicators and galaxy stellar mass measurements. In Section~\ref{sec:results} we present derived galaxy properties, and in Section~\ref{sec:discussion} we compare them to recent observations and simulations, discussing the implications of our results. Finally, we summarize our work in Section~\ref{sec:summary}. Throughout this work, we use ``ln" to represent the natural logarithm and ``log" to represent the base-10 logarithm. We also  assume a \cite{Chabrier2003} initial mass function (IMF) and adopt a flat $\Lambda$CDM cosmology with parameters from \cite{Planck2020} ($\Omega_m = 0.315$ and $H_0 = 67.4~\mathrm{km\,s^{-1}\,Mpc^{-1}}$).

\section{Data} \label{sec:data}

\subsection{Spectroscopy} \label{sec:spectroscopy}
We use \textit{JWST}/NIRSpec data from the Cosmic Evolution Early Release Science (CEERS) program (\textit{JWST} ERS\#1345, PI Finkelstein; \citealt{Finkelstein2017,Finkelstein2024}) and the Red Unknowns: Bright Infrared Extragalactic Survey (RUBIES) program (\textit{JWST} Cycle 2 GO\#4233, PIs de Graaff and Brammer; \citealt{Graaff2023,Wang2024,deGraaff2025}). 

CEERS observed six NIRSpec/PRISM MSA pointings in the Extended Groth Strip (EGS) field, each for an effective integration time of 3107s in the PRISM, G140M, G235M, and G395M disperers. We use the internal CEERS collaboration V0.9 NIRSpec reduction \citep[an evolution of the process detailed in ][]{ArrabelHaro2023ApJ,ArrabelHaro2023Nat}. The full details of these reductions will appear in P. Arrabal-Haro
et al. (2025, in preparation) and are summarized below. 

The NIRSpec data were reduced using the JWST Calibration Pipeline version 1.8.5 \citep{Bushouse2022_JWST_Pipeline} and CRDS context \texttt{jwst\_1029.pmap}. The data were reduced using the CRDS-prescribed pipeline parameters, with a few exceptions to improve the rejection of cosmic rays and `snowball' artifacts and to better optimize the width of the 2D to 1D spectral extraction apertures.

RUBIES observed six MSA pointings in the EGS field and 12 pointings in the UDS field, each for an effective exposure time of 2889s in both the PRISM and G395M dispersers. In this work, we use the reduction of this dataset first presented in \cite{Taylor2025}. Briefly, \citet{Taylor2025} adopted the same JWST Calibration Pipeline modifications used in the CEERS V0.9 reductions (see above), but used a newer pipeline version (1.13.4) and CRDS context (\texttt{1215.pmap}).

While this combined spectroscopic sample enables analysis of galaxies with a broad range of properties, it is subject to selection biases. Neither CEERS nor RUBIES implements a simple target selection function. CEERS targets are likely somewhat biased to bluer galaxies, as many were selected from pre-existing \textit{HST}-based photometry before \textit{JWST} imaging became available. RUBIES prioritized redder, brighter, \textit{JWST}-selected sources. We examine the burstiness parameter used in our analysis---the H$\alpha$-to-UV ratio---and find that its distribution does not significantly differ between the CEERS and RUBIES targets. This suggests that, despite differing selection strategies, the two surveys probe galaxies with comparable levels of recent star-formation variability. 

\subsection{Photometry} \label{sec:photometry}

We use \textit{JWST}/NIRCam imaging from the CEERS/EGS and PRIMER/UDS surveys.  For CEERS, we use the v1.0 reductions described in \citet{Bagley2023} and in the appendix of \citet{finkelstein2025}.  For PRIMER (PI Dunlop), we use an internal reduction from the PRIMER team (internal version 0.6). The PRIMER imaging data were reduced using the PRIMER Enhanced NIRCam Image Processing Library (PENCIL; Magee et al., in preparation, Dunlop et al. in preparation) software.  Both datasets include imaging in seven NIRCam broadbands (F090W, F115W, F150W, F200W, F277W, F356W, F444W) and one medium band (F410M).  Both datasets also include {\it HST}/ACS optical imaging in F606W and F814W, and WFC3/IR near-infrared imaging in F105W (partial coverage), F125W, F140W and F160W from the Cosmic Assembly Near-infrared Deep Extragalactic Legacy Survey (CANDELS; \citealt{Grogin2011,Koekemoer2011}), and partial coverage in F435W from UVCANDELS \citep{Mehta2024}.  Our photometric catalog follows the procedure outlined in \citet{Finkelstein2024}, with modest updates as described in several papers \citep[e.g.,][]{Taylor2025,kokorev2025}.  It is based on a combined F277W$+$F356W detection image, inclusive of PSF matching to ensure accurate colors across a wide wavelength baseline.  Typical flux limits are listed in Table 1 and 2 of \citet{Finkelstein2024}.

\section{Methods} \label{sec:methods}

\begin{deluxetable*}{cccc}[ht!]
\tablecaption{Summary of Derived Quantities}
\tablehead{Quantity & Method}

\startdata
redshift ($z$) & spectroscopic line identification\\
stellar mass ($M_*$) & SED fitting with \textsc{Bagpipes}\\
dust extinction ($A_V$) & spectral line fitting of H$\alpha$/H$\beta$ Balmer decrement \\
H$\alpha$ luminosity ($L_{H\alpha}$) & spectral line fitting of H$\alpha$, dust-corrected using $A_V$ values \\
UV luminosity ($L_{UV}$) & SED fitting with \textsc{Bagpipes}, dust-corrected using $A_V$ values
\label{tab:derived_quantities}
\enddata
\end{deluxetable*}

\begin{figure}[ht!]
    \centering
    \includegraphics[width=\linewidth]{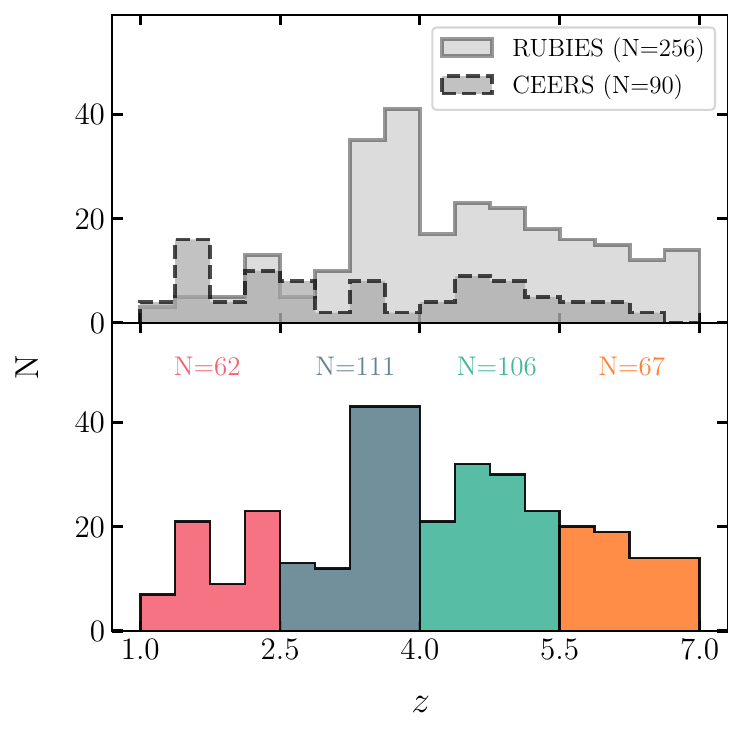}
    \caption{
    Spectroscopic redshift distribution of our sample split by survey (top) and by redshift bin (bottom). The full sample consists of 346 spectroscopically confirmed galaxies. We divide the sample into four redshift bins: $z=1$--$2.5$, $z=2.5$--$4$, $z=4$--$5.5$, and $z=5.5$--$7$, which we use throughout our analysis to examine redshift evolution in galaxy properties across cosmic time. 
    }
    \label{fig:sample_redshift_distribution}
\end{figure}

\subsection{Emission-Line Fitting} \label{sec:emission_line_fitting}

We obtain spectroscopic redshift and emission line information for 234 CEERS sources from an internal CEERS-team spectroscopic redshift compilation, predominantly based on strong emission lines. The emission line redshifts are measured using a Gaussian flux-weighted average of the centroids of the Gaussian profiles, based on the corresponding vacuum transition wavelengths from \citet{Hoof2018}. These measurements were performed using \textsc{LiMe} (see \citet{Fernandez2024}), with a visual inspection of the quality of the profile fittings. The line measurements from the CEERS sample are available online at https://ceers-data.streamlit.app. 

We obtain spectroscopic redshifts for \textit{JWST}/RUBIES sources from two methods. First, we collect spectroscopic redshifts available from the DAWN JWST Archive (DJA)\footnote{\url{https://s3.amazonaws.com/msaexp-nirspec/extractions/nirspec_graded_v3.html}}. We additionally supplement these redshifts with a bespoke emission line detection code. Briefly, this code searches for strong emission lines in NIRSpec spectra, and fits spectroscopic redshifts by comparing the velocity separations of detected lines to a list of strong emission lines typically observed in galaxies. This code is still a prototype, thus we visually inspect all spectroscopic redshifts estimated using this routine. In total, we compile 866 spectroscopic redshifts for RUBIES. 

We identify and remove 62 galaxies likely dominated by active galactic nuclei (AGN) using the sample of AGN-dominated sources from CEERS and RUBIES identified in \cite{Taylor2025}, which are selected based on the presence of broad Balmer emission lines. Additionally, we exclude sources at $z\geq7$ from the sample, due to the H$\alpha$ emission line detection limit with \textit{JWST}/NIRSpec. 

Spectral resolution varies depending on redshift, with G395M grating or PRISM spectra selected to optimize H$\alpha$ and H$\beta$ line detection. For sources at $4.9 \leq z < 6.9$, we use G395M grating spectra to measure both H$\alpha$ and H$\beta$ for dust correction via the Balmer decrement and to derive dust-corrected H$\alpha$ fluxes. For sources at $z < 3.4$ and $z \geq 6.9$, we use PRISM spectra for both purposes. For intermediate redshifts ($3.4 \leq z < 4.9$), we use PRISM spectra to measure the Balmer decrement and G395M spectra to derive dust-corrected H$\alpha$ fluxes.

For sources observed with NIRSpec G395M grating spectra, we model the blended [NII]+H$\alpha$ complex using a triple Gaussian profile and a linear continuum, allowing us to isolate the H$\alpha$ component by subtracting the [NII] contribution. The H$\beta$ emission line is modeled with a single Gaussian and a linear continuum. For sources observed with PRISM spectra, we model H$\alpha$ with a single Gaussian and a linear continuum due to the limited spectral resolution, and fit the blended H$\beta$+[OIII] complex using a triple Gaussian and a linear continuum, subtracting the [OIII] component to recover H$\beta$. Since we are unable to resolve the [NII] doublet with PRISM data, we correct the measured H$\alpha$+[NII] flux by applying a scaling factor of $0.98 \pm 0.02$, derived from the median H$\alpha$/([NII]+H$\alpha$) ratio measured in our sample of G395M sources, thereby estimating a deblended H$\alpha$ flux for PRISM sources.

We perform initial fits of the emission lines using the \textsc{curve\_fit} function from \textsc{SciPy} \citep{Virtanen2020} and pass these best-fit parameters as the starting points (with some perturbation to establish independent walkers) for the Markov-Chain-Monte-Carlo (MCMC) fitting using the \textsc{emcee} package for \textsc{Python} \citep{Foreman-Mackey2013}. We use 32 walkers and a maximum of 3000 steps to sample the likelihood function and reach convergence, always discarding the first 2000 burn-in samples before evaluating the best-fit parameters. 

We impose uniform priors on all model parameters: (1) emission line amplitudes are constrained to be positive and less than twice the best-fit value returned by \textsc{curve\_fit}, in order to guide convergence while still allowing flexibility in the MCMC optimization; (2) line centers are allowed to vary within $\pm 20~\mathring{A}$ of the expected central wavelength; (3) Gaussian widths are limited to the range $1.5$–$30~\mathring{\mathrm{A}}$ for the G395M grating and $30$–$200~\mathring{\mathrm{A}}$ for the PRISM data. These limits reflect the instrumental resolution floor of each mode, account for moderate emission line blending, and accommodate typical galaxy velocity dispersions up to a few hundred $\text{km} \cdot \text{s}^{-1}$; and (4) linear continuums are required to be positive. For blended line complexes, we also enforce fixed [NII] and [OIII] doublet ratios of 2.94:1 and 2.98:1, respectively, between the redder and bluer components. For each parameter, we report the median value of the MCMC sample distribution and we calculate the 1$\sigma$ confidence interval by reporting the 16th and 84th percentile values. Calculating the area under each Gaussian fit, we obtain flux measurements of the emission lines present in each galaxy. 

Through visual inspection of the line fits, we find and remove 5 broad-line AGN. We restrict our sample to galaxies with a signal-to-noise ratio $\ge3$ for both the H$\alpha$ and H$\beta$ emission lines, such that we can compute a robust dust correction via the Balmer decrement. This results in 91/234 CEERS and 257/866 RUBIES star-forming galaxies with strong H$\alpha$ and H$\beta$ emission line detections. Of these 348 sources, the Balmer decrement and H$\alpha$ emission was measured with the G395M grating and PRISM for 133 and 94 sources, respectively. For the remaining 121 sources, the Balmer decrement was measured with PRISM and H$\alpha$ emission was measured with the G395M grating. 

We assume a \cite{Calzetti2001} dust law to correct for nebular attenuation, with $\rm{}E(B-V)$ values derived from the Balmer decrement. We adopt an intrinsic ratio of 2.87 for $\rm{}H\alpha/H\beta$, corresponding to Case B recombination at an electron temperature $T_e=10,000$~K and electron density $n_e = 10^2$~cm$^{-3}$ \citep{Osterbrock1989}. Finally, we convert the dust-corrected emission line fluxes to luminosities using the spectroscopic redshift of each galaxy.

While the Case B recombination assumption is commonly used, we note that it may not hold for all galaxies---particularly in extreme ionization or density conditions---which can introduce deviations from the assumed intrinsic ratio and yield unphysical reddening values (i.e., $\mathrm{E(B{-}V)} < 0$) \citep[e.g.,][]{McClymont2025-2}. We find 119/348 galaxies in our sample exhibit such negative values (ranging $-0.9 \leq \mathrm{E(B{-}V)} < 0$) and we enforce $\mathrm{E(B{-}V)}=0$ for these cases to ensure physically meaningful results. The resulting distribution of extinction values spans $A_V = 0.0$–$5.3$~mag, with a median of $A_V = 0.5^{+1.2}_{-0.5}$~mag. We acknowledge that this non-negligible population of sources with forced zero reddening may impact the scatter in derived quantities that depend on dust correction, such as the H$\alpha$-to-UV ratio. Additionally, we emphasize that the true attenuation law may vary among galaxies and depend on galaxy properties such as dust geometry and optical depth \citep[e.g.,][]{Tacchella2022-1,Tacchella2022-2,Reddy2025}.  

\subsection{SED Fitting} \label{sec:SED_fitting} 

In this section we outline how we fit the spectral energy distribution (SED) of each galaxy using the full suite of available photometry from \textit{JWST}/NIRCam, \textit{HST}/WFC3, and \textit{HST}/ACS, spanning the rest-frame UV to IR. We cross-match the spectroscopic sample with the PRIMER/UDS, and CEERS, CANDELS/EGS photometric catalogs. We require the photometric and spectroscopic source positions to be separated by $<0.3$~arcsec. Some sources in the EGS field do not have NIRCam coverage; for those we use HST photometry from CANDELS. 

We perform stellar population modeling with the SED fitting code \textsc{Bagpipes} \citep{carnall18,carnall19}, to measure the stellar masses and UV-continuum magnitude (M$_{1500}$) from the posterior spectra. We adopt a non-parametric star-formation history (SFH) with the `bursty continuity' prior from \citet{tacchella23} \citep[also see][for more discussion regarding burstiness priors]{leja19}. Our SFH model used logarithmically spaced time bins defined in lookback time. 

The Universe's age at the redshift of interest is converted to Myr, and SFH bins are defined from 10~Myr up to just before the cosmic age using 10 logarithmic intervals (i.e. [10, 50, 100, ..., $t_{universe}$(z)]). An initial bin edge at 0~Myr is added at the beginning of the bin edge to ensure the full lookback time is covered.

We apply the Calzetti dust attenuation law \citep{Calzetti2000}, allowing the $A_V$ parameter to vary between 0 and 3 magnitudes. Nebular emission is included, with the ionization parameter ($\log U$) allowed to vary between $-4$ and $-1$.

For the SFH, 
we specify the total mass formed, metallicity, and fractional deviations in SFR ($\Delta \log\,\mathrm{SFR}$) in each time bin. We allow:
\begin{itemize}
    \item the total stellar mass formed to range between $10^{-4}$ and $10^{13}$~M$_\odot$,
    \item metallicity between 0.01 and 5~Z$_\odot$, using a logarithmic prior (i.e., log$_{10}$ prior).
\end{itemize}

Each SFH bin’s star formation deviation parameter is assigned a Student’s $t$-distribution prior with a prior scale of 2.0 to allow for bursty behavior, whenever present. These priors allow the model to flexibly fit a wide range of galaxy SFHs, including strongly burst-dominated scenarios.

The observed spectral energy distribution (SED) is interpolated to obtain the monochromatic flux density at 1500\,\AA\ in the rest frame ($F_{\nu,1500}$), which is then converted to an absolute AB magnitude at 1500\,\AA ($M_{\text{UV}}$). 

To characterize the stellar masses of our NIRSpec sample, we compare them to those of a large ($N=101,355$) CEERS photometric sample derived with the \textsc{dense basis} SED-fitting code \citep{Iyer2017,Iyer2019}, as described in \cite{Chworowsky2024}. For each NIRSpec galaxy at spectroscopic redshift $z_i$, we identify a subsample ($N=3000$) of CEERS galaxies with photometric redshifts closest to $z_i$ and compute the CEERS conditional stellar-mass percentiles at that redshift (16th–84th and 5th–95th). We find that 67.9\% of our NIRSpec galaxies fall within the middle 68\% of CEERS stellar masses at a matched redshift, and 93.2\% fall within the middle 90\%. These coverages are consistent with the nominal values expected for a random subsample, indicating that our NIRSpec sample fairly traces the CEERS stellar-mass–redshift distribution.

\subsection{UV-Continuum Measurement} \label{sec:UV_continuum_emission_measurement}

We convert the UV-continuum absolute magnitudes ($M_{\text{UV}}$) into luminosities ($L_{\text{UV}}$). To dust-correct these measurements, we apply the $\rm E(B-V)$ values derived from the Balmer decrement, as described in Section~\ref{sec:emission_line_fitting}. To account for the differential attenuation between nebular emission lines and stellar continuum light, we scale the nebular $\rm E(B-V)$ values by a factor of 0.44 \citep{Calzetti2000}. We require a signal-to-noise ratio $\ge 3$ for the dust-corrected UV luminosities, which reduces the sample size by two galaxies.

The final combined CEERS and RUBIES sample includes 346 galaxies at $0.8 \leq z < 7$, with stellar masses spanning $7.0 \leq \text{log}(M_* / M_{\odot}) \leq 10.9$. We summarize the physical quantities measured and derived above in Table~\ref{tab:derived_quantities} and show the spectroscopic redshift distribution of this sample in Figure~\ref{fig:sample_redshift_distribution}. We separate the sample into redshift bins of $z=1$--$2.5$, $z=2.5$--$4$, $z=4$--$5.5$, and $z=5.5$--$7$ to track changes in galaxy properties with cosmic time. These boundaries were chosen to ensure a minimum of $\sim60$ galaxies per bin, enabling statistically robust population-level analyses while also aligning with key epochs in cosmic star formation history.

\subsection{Stellar-Mass Completeness} \label{sec:stellar-mass-completeness}

Before characterizing the SFMS, we construct a stellar-mass complete sample. We estimate stellar–mass completeness in each redshift bin following the method described in Section~5.2 of \cite{Pozzetti2010}. For each galaxy, we compute the limiting stellar mass ($M_{\text{lim}}$) it would have if its apparent magnitude ($m_i$) were equal to the limiting magnitude of the survey ($m_{\text{lim}}$), assuming the galaxy has the same mass–to–light ratio. We define $m_{\text{lim}}$ using the NIRCam F277W/F356W bands, where the median catalog $5\sigma$ depth is $m_{\text{lim}}=28.5$ \citep{finkelstein2025}. These filters represent the continuum sensitivity that drives our stellar-mass selection. For each galaxy with stellar mass $M_{*,i}$, the limiting stellar mass is given by
\begin{equation}
\log (M_{\rm lim,i}) = \log (M_{*,i}) +0.4\,(m_i-m_{\rm lim}) \quad .
\label{eq:mass_completeness}
\end{equation}

In each redshift bin, we evaluate $M_{\rm lim,i}$ for the faintest 30\% of galaxies and take the 90th percentile as the mass–completeness threshold. Our mass-completeness limits are shown in Table~\ref{tab:mass_completeness}. Galaxies below these thresholds are excluded from subsequent SFMS fitting.

\begin{deluxetable}{cccc}[ht!]
\tablecaption{Stellar-Mass Complete Sample Parameters}
\tablehead{Redshift Bin & $N$ & $\log(M_{\text{lim}}/M_{\odot})$ & $\log(M_{\text{piv}}/M_{\odot})$ }

\startdata
$1 \lesssim z \leq 2.5$ & $37$ & $8.4$ & $9.0$ \\
$2.5 < z \leq 4$ & $88$ & $8.2$ & $9.1$ \\
$4 < z \leq 5.5$ & $42$ & $8.7$ & $9.3$ \\
$5.5 < z < 7$ & $19$ & $8.7$ & $9.5$ \\
\enddata
\tablecomments{For each redshift bin, $M_{\text{lim}}$ is the limiting stellar mass computed in Section~\ref{sec:stellar-mass-completeness} and $M_{\text{piv}}$ is the median stellar mass used for SFMS fitting in Section~\ref{sec:the_star_forming_main_sequence}.}
\label{tab:mass_completeness}
\end{deluxetable}

\section{Results} \label{sec:results}

\subsection{The Star-Forming Main Sequence} \label{sec:the_star_forming_main_sequence}

We calculate dust-corrected H$\alpha$- and UV-continuum-based SFRs using the calibrations from \citet{Kennicutt1998}, and apply a $-0.24$~dex correction to convert from a \citet{Salpeter1955} to a \citet{Chabrier2003} IMF. 

We model the relationship between SFR and stellar mass ($M_*$) using a linear relation of the form
\begin{equation}
y = m \times (\log{M_*} - \log{M_{\text{piv}}}) + b \quad ,
\label{eq:SFMS_linear_model}
\end{equation}
where $y = \log(\text{SFR}/M_\odot~\text{yr}^{-1})$, $m$ is the slope, and $b$ is the normalization at the pivot mass $M_{\text{piv}}$. We define $M_{\text{piv}}$ for each redshift bin in Table~\ref{tab:mass_completeness} as the median stellar mass. This reduces covariance between the slope and intercept during fitting.

To model observed scatter around the relation, we assume each observed data point $y_{\text{obs}, i}$ is drawn from a Gaussian distribution centered on the model prediction $y_{\text{model}, i}$ with total variance
\begin{equation}
y_{\text{obs},i} \sim \mathcal{N}(y_{\text{model},i},\,\sigma^{2}_{i}) \quad ,
\label{eq:normal_dist}
\end{equation}
\begin{equation}
\sigma_i^2 = \sigma_{\text{obs}, i}^2 + \sigma_{\text{int}}^2 \quad ,
\label{eq:total_variance}
\end{equation}
where $\sigma_{\text{obs}, i}$ is the $1\sigma$ measurement uncertainty on the $\log(\text{SFR})$ of the $i$-th data point, computed by propagating the SFR measurement error into log space, and $\sigma_{\text{int}}$ represents the contribution of intrinsic scatter in the SFMS relation to the spread in $\log(\text{SFR})$.

To account for sensitivity limits that impact the detectability of galaxies with faint star formation activity, we estimate limiting SFR values separately for H$\alpha$ and UV-continuum observations. For our H$\alpha$-based measurements, this limit is set by the sensitivity of the NIRSpec G395M grating, with a 5$\sigma$ detection threshold of $1.1\times 10^{-18}~\text{erg}~\text{s}^{-1}~\text{cm}^{-2}$ \citep{finkelstein2025}. Although some H$\alpha$-based SFRs were measured with PRISM, the higher spectral resolution of G395M defines an upper limit on our detection capabilities for this sample. For the UV-continuum-based measurements, we estimate a 3$\sigma$ limiting SFR(UV) using the posterior distributions of rest-frame UV-continuum magnitudes ($M_{\text{UV}}$) from our SED-fitting (Section~\ref{sec:SED_fitting}). For each redshift bin, we adopt the maximum of the 16th percentile $M_{\text{UV}}$ values across the sample, representing the faintest reliably measured UV-continuum level consistent with the photometry. 

To model completeness effects near these limits, we incorporate a sigmoid-based selection function $\Phi(y)$:
\begin{equation}
\Phi(y) = \frac{1}{1 + \exp[-\alpha \times (y - \log_{10}(\text{SFR}_{\text{lim}}))]} \quad ,
\label{eq:selection_function}
\end{equation}
where $\text{SFR}_{\text{lim}}$ is the limiting SFR and $\alpha$ controls the sharpness of the transition from incomplete to complete. We fix $\alpha = 5$ for all redshift bins, corresponding to a moderate transition in completeness over a range of $\sim$0.3 dex. This formulation allows the model to reduce the weight of galaxies near the sensitivity threshold without discarding them, mitigating bias in the slope and scatter estimates.

We model the log-likelihood function for our fit as
\begin{equation}
\ln \mathcal{L}(\theta) = -\frac{1}{2} \sum_i \left[ \frac{(y_{\text{obs}, i} - y_{\text{model}, i})^2}{\sigma_i^2} + \ln (\sigma_i^2) \right] +
\sum_i \ln \Phi(y_{\text{obs},i}) \quad ,
\label{eq:log_likelihood_selection}
\end{equation}
where $\theta = \{m,\, b,\, \sigma_{\text{int}}\}$ are the model parameters.

\begin{figure*}

    \centering
    \includegraphics[width=0.49\textwidth]{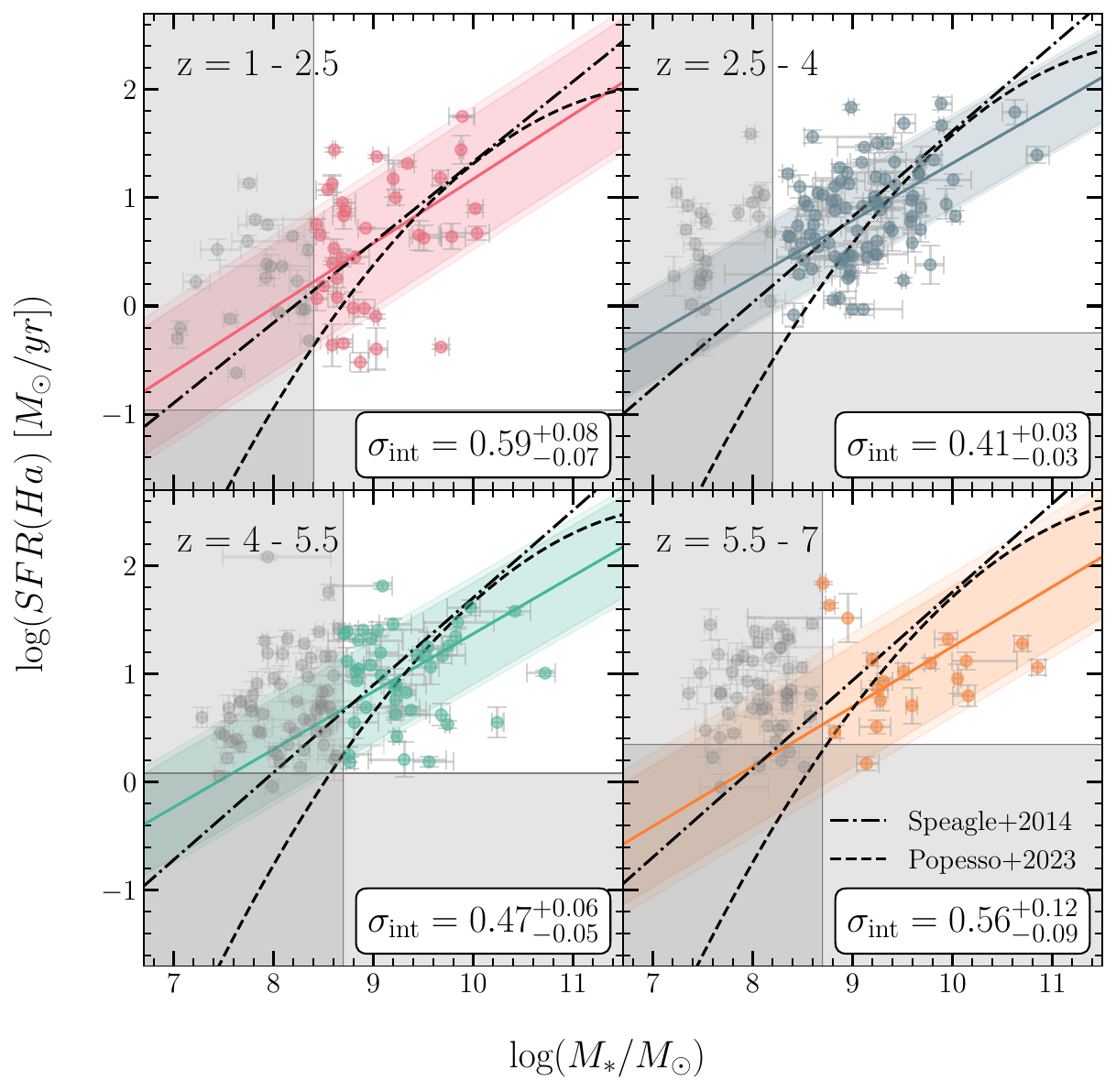}
    \includegraphics[width=0.49\textwidth]{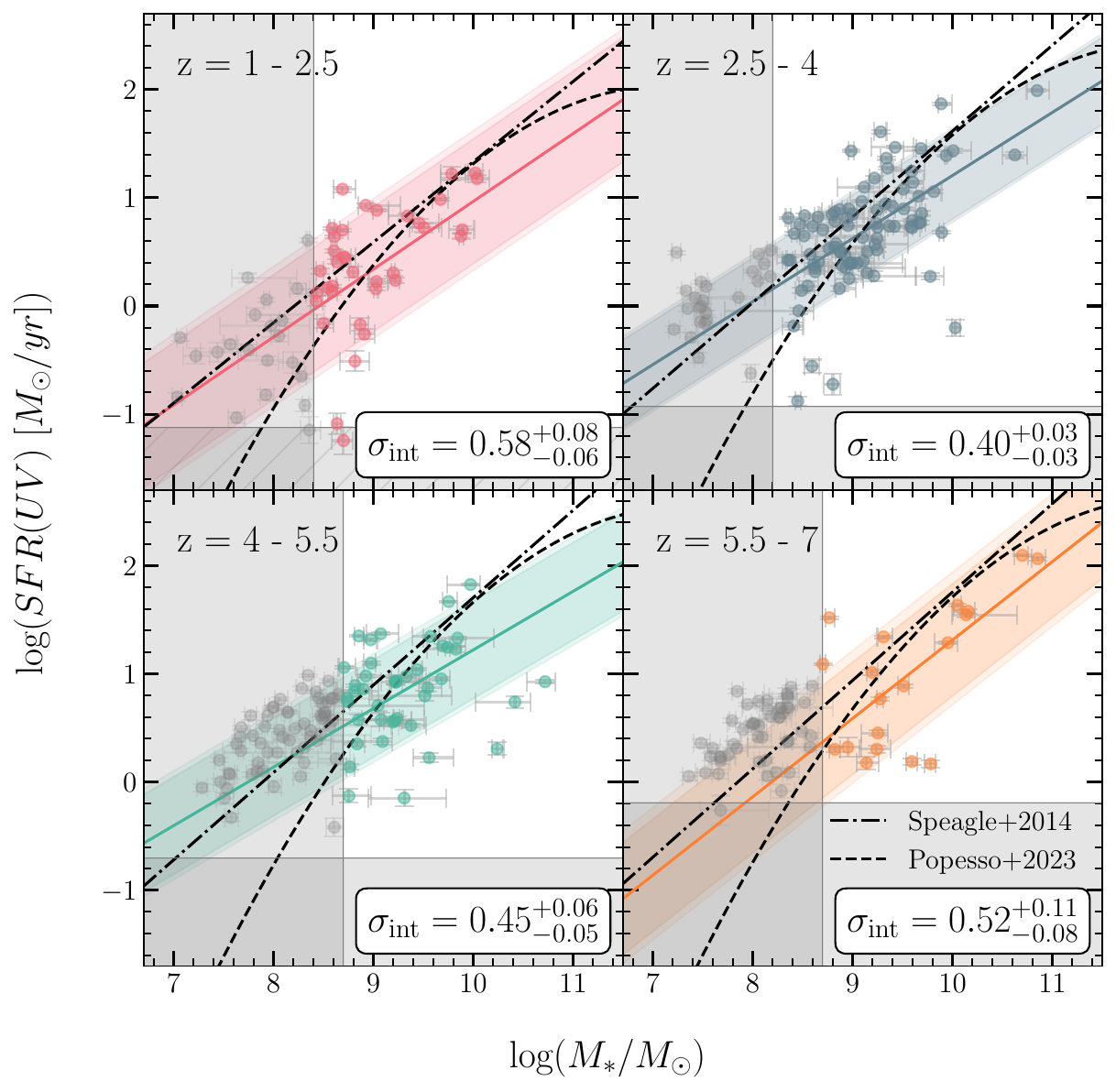}
    
    \caption{H$\alpha$- (left) and UV-continuum-based (right) SFRs vs. stellar masses for our galaxy sample binned by redshift. Vertical gray regions and gray data points indicate where our sample is not mass complete, as defined in Section~\ref{sec:stellar-mass-completeness}. These data points are excluded from the fitting procedure. In the left panel, horizontal gray regions indicate the 5$\sigma$ NIRSpec G395M grating emission line sensitivity limits derived from Table~6 in \cite{finkelstein2025}. In the right panel, we compute 3$\sigma$ limiting SFR(UV) values based on the faintest reliably measured rest-frame UV-continuum in each redshift bin, defined as the maximum of the 16th percentile $M_{1500~\mathring{A}}$ values from our SED-fitting posteriors of galaxies with $\text{SNR}>3$. The colored shaded regions denote the 1$\sigma_{\rm{}int}$ intrinsic scatter intervals about the SFMS and its associated error. The black dashed and dashed-dotted curves represent literature values from \cite{Popesso2023} and \cite{Speagle2014}, respectively. On average, the H$\alpha$-based SFMS and the UV-based SFMS exhibit the same degree of intrinsic scatter, indicating that we do not detect short-timescale ($\sim10$~Myr) SFR variability across the sample with the SFMS.}
    
    \label{fig:SFMS_Ha_and_UV}
\end{figure*}

We estimate the posterior distributions of the parameters using an MCMC sampler with 20 walkers and 5000 total steps, discarding the first 3000 steps as burn-in. We adopt uniform priors: $m \in [0.5, 1.5]$, $b \in [-5, 5]$, and $\sigma_{\text{int}} \in [0, 1.5]$, consistent with constraints from prior studies \citep[e.g.,][]{Speagle2014,Whitaker2014,Salmon2015,Popesso2023}. We report the median and 16th–84th percentile values from the posterior distributions. We plot the H$\alpha$- and UV-continuum-based SFRs versus stellar masses in Figure~\ref{fig:SFMS_Ha_and_UV} and provide our best-fit results in Table~\ref{tab:SFMS_parameters}. Horizontal gray regions indicate areas where we do not expect to detect galaxies due to survey sensitivity limits. The regime where our sample is mass-incomplete, as defined in Section~\ref{sec:stellar-mass-completeness}, is indicated by the vertical gray regions and gray data points.

We compute the intrinsic scatter---averaged across the probed redshift range---for both SFMS relations to be ${\langle \sigma_{\rm{}H\alpha} \rangle}=0.45 \pm 0.03$~dex and ${\langle \sigma_{\rm{}UV} \rangle}=0.44 \pm 0.02$~dex. The difference, ${\langle \sigma_{\rm{}H\alpha} \rangle} - {\langle \sigma_{\rm{}UV} \rangle} = 0.01 \pm 0.04$~dex, is consistent with zero and therefore not statistically significant for our sample. 

\begin{deluxetable*}{cccccc}[ht!]
\tablecaption{Star-forming Main Sequence Best-fit Parameters}
\tablehead{SFR Indicator & Redshift Bin & $m$ & $b$ & $\sigma_{\text{int}}$ }

\startdata
H$\alpha$ & $1 \lesssim z \leq 2.5$ & $0.59^{+0.12}_{-0.07}$ & $0.58^{+0.10}_{-0.09}$ & $0.59^{+0.08}_{-0.07}$ \\
H$\alpha$ & $2.5 < z \leq 4$ & $0.53^{+0.04}_{-0.02}$ & $0.85^{+0.05}_{-0.05}$ & $0.41^{+0.03}_{-0.03}$ \\
H$\alpha$ & $4 < z \leq 5.5$ & $0.53^{+0.06}_{-0.03}$ & $1.00^{+0.07}_{-0.07}$ & $0.47^{+0.06}_{-0.05}$ \\
H$\alpha$ & $5.5 < z < 7$ & $0.55^{+0.09}_{-0.04}$ & $1.00^{+0.13}_{-0.13}$ & $0.56^{+0.12}_{-0.09}$ \\
\hline
UV & $1 \lesssim z \leq 2.5$ & $0.63^{+0.15}_{-0.09}$ & $0.34^{+0.10}_{-0.10}$ & $0.58^{+0.08}_{-0.06}$ \\
UV & $2.5 < z \leq 4$ & $0.58^{+0.08}_{-0.05}$ & $0.69^{+0.04}_{-0.04}$ & $0.40^{+0.03}_{-0.03}$ \\
UV & $4 < z \leq 5.5$ & $0.54^{+0.06}_{-0.03}$ & $0.84^{+0.07}_{-0.07}$ & $0.45^{+0.06}_{-0.05}$ \\
UV & $5.5 < z < 7$ & $0.72^{+0.18}_{-0.14}$ & $0.98^{+0.12}_{-0.12}$ & $0.52^{+0.11}_{-0.08}$ \\
\enddata

\tablecomments{All parameters are computed from galaxy SFRs as a function of stellar mass. For each redshift bin, parameters are derived within the range that we are stellar-mass complete, as outlined in Table~\ref{tab:mass_completeness}. Note that $b$ is the normalization at the subsample median stellar mass ($M_{\text{piv}}$).}
\label{tab:SFMS_parameters}
\end{deluxetable*}

The mass-incomplete subsamples exhibit visually flatter H$\alpha$-based SFMS relations compared to the best-fit relations given by the mass-complete subsample and relations reported by \citet{Popesso2023} and \citet{Speagle2014}, especially at high redshifts ($z \gtrsim 5.5$). This deviation likely arises from two factors: (1) sensitivity limits in the NIRSpec observations that bias against low-SFR galaxies, artificially elevating the observed SFMS, and (2) increased star-formation variability in low-mass galaxies at early times, enhancing the number of galaxies observed during a burst phase \citep[e.g.,][]{Sparre2017,Pelliccia2020,Sun2023a,Asada2024,Daikuhara2024,McClymont2025-1,Gelli2025}. Together, these effects limit the SFMS's ability to recover the full distribution of star-formation states and may obscure burst-driven variability in the H$\alpha$ scatter.

\subsection{The H$\alpha$-to-UV Luminosity Ratio} \label{sec:Ha_to_UV_luminosity_ratio}

To more directly probe burstiness in galaxy SFHs, we turn to an independent diagnostic: the H$\alpha$-to-UV luminosity ratio. In our analysis, we use attenuation corrected luminosities for both H$\alpha$ and UV. This ratio is sensitive to recent changes in SFR on timescales of $\sim$10–100~Myr and offers a complementary view to the SFMS-based analysis. 

\begin{figure}[ht!]
    \centering
    \includegraphics[width=\linewidth]{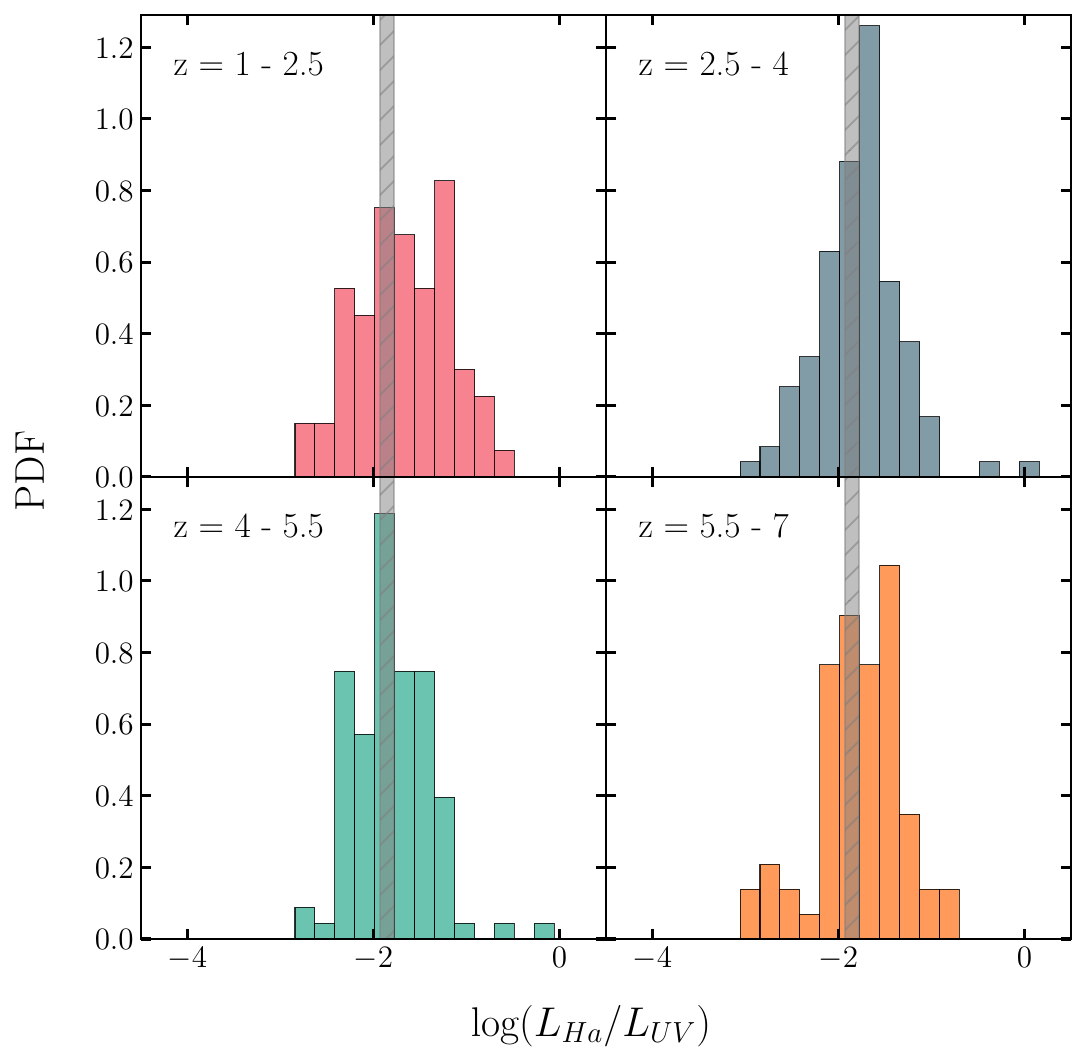}
    \caption{The distribution of H$\alpha$-to-UV ratio values binned by redshift. The vertical gray hatched region shows the equilibrium range of the H$\alpha$-to-UV ratio calculated by \cite{Mehta2023}. For each redshift bin, the distribution peak lies within the equilibrium range expected for a constant SFH over the past $\sim 100$~Myr.}
    \label{fig:lum_ratio_distribution}
\end{figure}

We explore how the H$\alpha$-to-UV luminosity ratio of galaxies in our sample changes with redshift and stellar mass. We compare these results to an equilibrium range of log$(L_{\text{H}\alpha}/L_{\text{UV}})=[-1.93, \, -1.78]$ as predicted by stellar population synthesis models assuming a constant SFH over the past $\sim 100$~Myr and metallicities (log($Z/Z_{\odot}$)) between $-2$ and $0$ \citep{Mehta2023}. This convention is widely adopted in the literature \citep[e.g.,][]{Glazebrook1999,Sullivan2000,Sullivan2001,Broussard2019,Emami2019,Faisst2019,Atek2022,Broussard2022,Asada2024,Endsley2024,Clarke2024} as it reflects the characteristic timescales traced by H$\alpha$ and UV emission. Thus, the equilibrium range provides a consistent benchmark for identifying short-timescale burstiness independent of cosmic epoch. In Figure~\ref{fig:lum_ratio_distribution}, we show the H$\alpha$-to-UV ratio distributions for each redshift bin. The distributions are approximately Gaussian and peak within the equilibrium range expected for a constant SFH over the past $\sim100$~Myr. However, the distribution at $z=5.5$--$7$ exhibits a tail or possible bimodality toward lower H$\alpha$-to-UV ratio values. We interpret this feature as originating from high-mass ($9\lesssim\text{log}(M_*/M_{\odot})\leq10.9$) galaxies that fall below the H$\alpha$-based SFMS in Figure~\ref{fig:SFMS_Ha_and_UV}. This suggest a recent decline in star formation activity among massive galaxies at high redshift.

\begin{figure*}
    \centering
    \includegraphics[width=\textwidth]{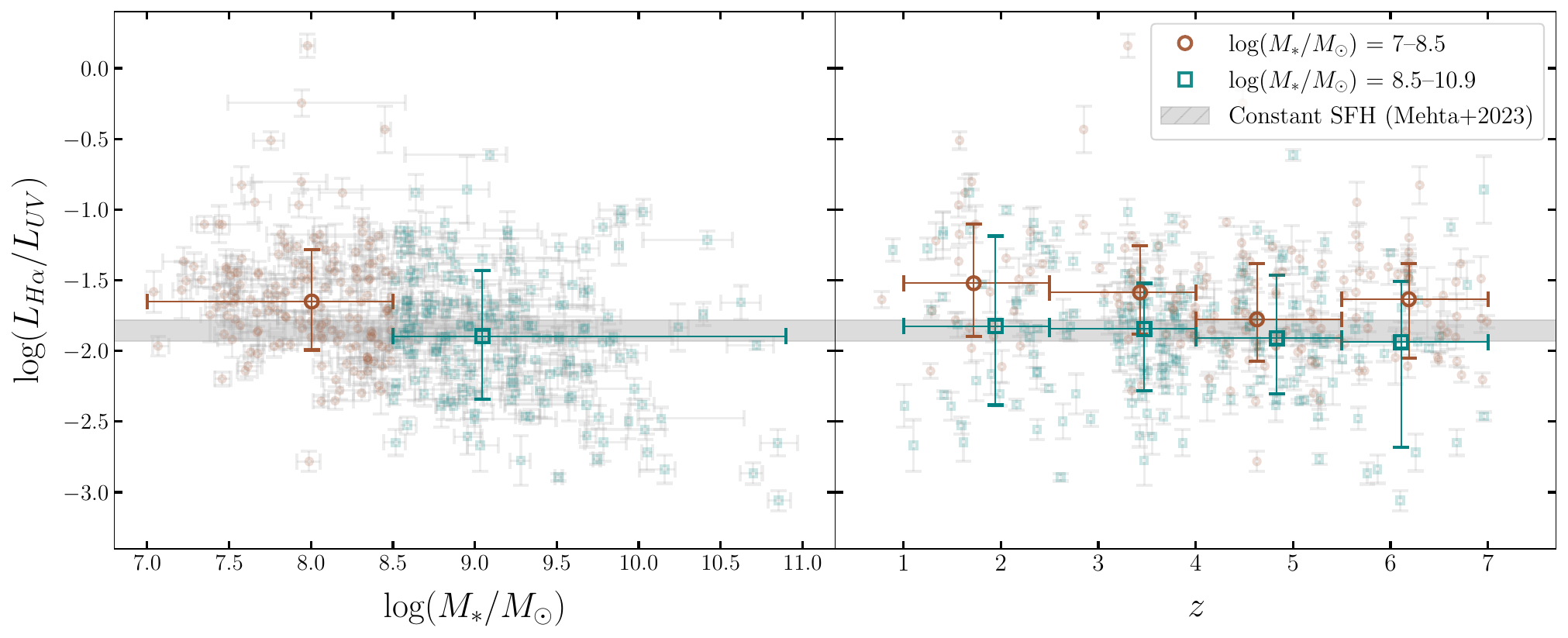}
    \caption{Left panel: The H$\alpha$-to-UV ratio vs. stellar mass for our galaxy sample binned by stellar mass. The large points represent the median values in each stellar mass bin. The vertical error bars denote the 16th and 84th percentiles of each bin and the horizontal error bars denote the stellar mass range of each bin. The horizontal gray hatched region shows the equilibrium range of the H$\alpha$-to-UV ratio calculated by \cite{Mehta2023}. Right panel: The H$\alpha$-to-UV ratio vs. redshift for our galaxy sample binned by stellar mass. The large points represent the median values in each stellar mass and redshift bin. The vertical error bars denote the 16th and 84th percentiles of each bin and the horizontal error bars denote the redshift range of each bin. Lower-mass galaxies maintain higher H$\alpha$-to-UV ratio values across all redshifts probed compared to higher-mass galaxies, suggesting lower-mass galaxies are more prone to bursty star formation.}
    \label{fig:lum_ratio_redshift_mass}
\end{figure*}

In Figure~\ref{fig:lum_ratio_redshift_mass}, we plot the H$\alpha$-to-UV ratio as a function of stellar mass and redshift. For each stellar mass bin, we quantify the scatter in the H$\alpha$-to-UV ratio using the standard deviation and estimate its uncertainty via bootstrap resampling. We find no significant difference in the average scatter between lower-mass ($7\leq\text{log}(M_*/M_{\odot})<8.5$) and higher-mass ($8.5\leq\text{log}(M_*/M_{\odot})\leq10.9$) galaxies in our sample, with $\langle \sigma_{\text{low-mass}} \rangle - \langle \sigma_{\text{high-mass}} \rangle = -0.04 \pm 0.04$~dex (left panel). However, across all redshifts, lower-mass galaxies exhibit higher H$\alpha$-to-UV ratios, consistent with more burst-driven star formation (right panel). 

\section{Discussion} \label{sec:discussion}

\subsection{Fraction Around the H$\alpha$-to-UV Equilibrium} \label{sec:frac_around_eq}

To characterize these results further, we compute the fraction of galaxies in our sample that lie above ($f_{\text{above}}$), below ($f_{\text{below}}$), and within ($f_{\text{eq}}$) the H$\alpha$-to-UV ratio equilibrium range calculated by \cite{Mehta2023}, which corresponds to a constant SFH over the past $\sim100$~Myr. A galaxy is classified as within the equilibrium range if its 1$\sigma$ uncertainty interval overlaps the range defined by \citet{Mehta2023}.

We compute these fractions using a probabilistic approach that propagates individual measurement uncertainties into the population-level distribution. For each galaxy, we perturb the observed H$\alpha$-to-UV ratio by drawing from a Gaussian distribution centered on the measured value with a standard deviation set by its 1$\sigma$ uncertainty. We then resample these perturbed values via bootstrap resampling and compute the fractions for each resample. The final estimates for $f_{\text{above}}$, $f_{\text{below}}$, and $f_{\text{eq}}$ are taken as the median of the bootstrap distributions, and the uncertainties are defined by the 16th and 84th percentiles. 

We find that $73^{+4}_{-4}$~\% of galaxies in our sample have H$\alpha$-to-UV ratios that lie either above or below the equilibrium range defined by \cite{Mehta2023}, and are therefore inconsistent with a constant SFH over the past $\sim$100~Myr. We find no variation in this statistic between our lower-mass and higher-mass samples. Furthermore, averaging across all redshifts, we find lower-mass galaxies exhibit a higher fraction above $f_{\text{above}}= 0.53^{+0.05}_{-0.05}$ and a lower fraction below $f_{\text{below}} = 0.18^{+0.04}_{-0.03}$ compared to higher-mass galaxies with $f_{\text{above}}= 0.33^{+0.04}_{-0.04}$ and $f_{\text{below}} = 0.41^{+0.04}_{-0.04}$.

For comparison, \citet{Clarke2024} report that $48-71$\% of their spectroscopic sample of 146 galaxies from CEERS and JADES at $z \sim 1.4$--$7$ and with $7.0 \leq \text{log}(M_* / M_{\odot}) \leq 10.5$ deviate from a constant SFH. Applying the same selection to our sample, we recover ${73}^{+4}_{-4}$~\%, which is consistent within uncertainties. At $z \sim 4.7$--$6.5$ and with $6.0 \leq \text{log}(M_* / M_{\odot}) \leq 10.5$, \citet{Asada2024} find 60\% of their photometric sample of 125 galaxies from CANUCS deviate from a constant SFH. For a matched subsample, we measure ${72}^{+4}_{-4}$~\%, a modestly higher fraction that may reflect differences in sample selection and methodology (e.g., SED-derived galaxy properties from photometry and inclusion of gravitationally lensed systems).

In their analysis of 368 galaxies at $z\sim6$ with $-22.0 \leq M_{\text{UV}} \leq -15.8$, \cite{Endsley2024} found that high-redshift UV-faint galaxies may be more likely to be observed in recent downturns in their SFR, and therefore have H$\alpha$-to-UV ratios below the equilibrium range. These galaxies, which exhibit suppressed H$\alpha$ emission relative to their UV continuum, are often referred to as ``temporarily quiescent"---systems that have recently experienced a decline in SFR following a burst phase, but may resume elevated star formation on short timescales. Similarly, the SERRA zoom-in simulations \citep{Gelli2025} suggest that the fraction of temporarily quiescent galaxies at $6<z<8$ increases with decreasing mass (from $\lesssim 40$\% at $M_* > 10^8 M_{\odot}$ to $\gtrsim 60$\% at $M_* < 10^8 M_{\odot}$). Additional simulation-based work using IllustrisTNG and ASTRID indicates that lower-mass quenched galaxies tend to reside in denser environments than their star-forming counterparts \citep{Weller2025}, pointing to environment as a potential driver of quenching at high redshift. 

In our sample, fractional measurements for galaxies with similar properties to these works (high redshift: $5.5 \leq z < 7$, low stellar mass: $7.0 \leq \text{log}(M_* / M_{\odot}) < 8.5$, and faint: $-20.83 \leq M_{\text{UV}} \leq -18.16$) are given in the second-to-last row in Table~\ref{tab:frac_vals}. We find a high fraction above the equilibrium range ($f_{\text{above}} = 0.53^{+0.06}_{-0.09}$) and a low fraction below it ($f_{\text{below}} = 0.19^{+0.06}_{-0.06}$), suggesting that a substantial portion of our lower-mass, UV-faint galaxies are observed during burst phases. These empirical results contrast with the trends inferred from some simulations and observational studies that predict a higher prevalence of temporarily quiescent systems in this regime. 

\begin{figure*}
    \centering
    \includegraphics[width=\textwidth]{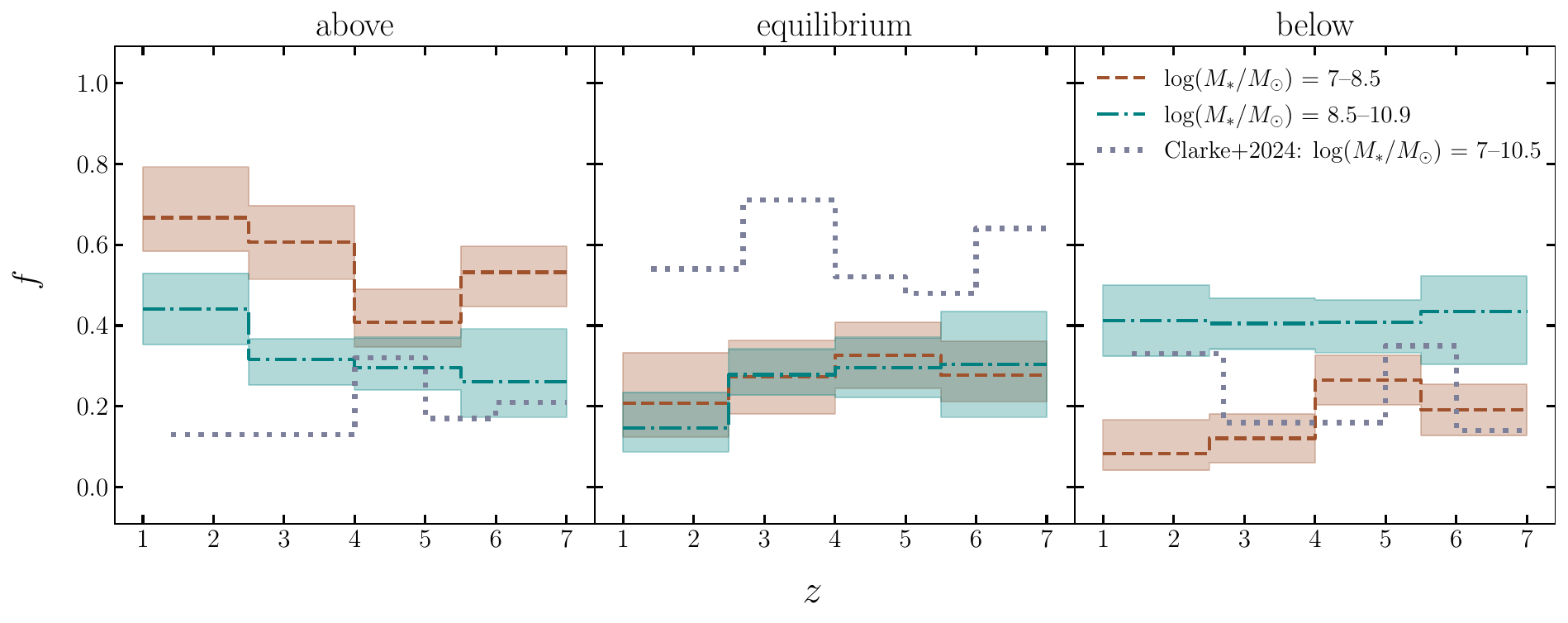}
    \caption{The fraction of galaxies whose H$\alpha$-to-UV ratios are consistent (middle panel) and inconsistent (outer panels) with a constant SFH computed by \cite{Mehta2023} shown as a function of redshift. For each stellar mass and redshift bin, we perturb the observed ratios using Gaussian noise based on 1$\sigma$ errors and compute fractions via bootstrap resampling. Measured values correspond to the median of the bootstrap distribution, with error bars from the 16th and 84th percentiles. Colored curves represent this work, binned by stellar mass and redshift, compared to the dotted gray curve from \cite{Clarke2024}, which includes galaxies with stellar masses in the range $10^7$--$10^{10.5}~M_*/M_{\odot}$. Within each bin, we compute the median redshift value. We estimate uncertainties on the fractions using bootstrap resampling. These trends suggest little evolution in the relative fraction of galaxies above, below, or within the equilibrium range over $z=1$--$7$.}
    \label{fig:lum_ratio_fraction_around_equilibrium}
\end{figure*}

\begin{deluxetable*}{ccccc}[ht!]
\tablecaption{Fraction of Galaxies Around the H$\alpha$-to-UV Ratio Equilibrium Range \label{tab:frac_vals}}
\tablehead{Redshift Bin & Mass Bin & $f_{\rm above}$ & $f_{\rm eq}$ & $f_{\rm below}$}
\startdata
$1 \lesssim z \leq 2.5$ & $7<\text{log}(M_*/M{\odot})<8.5$ & $0.67^{+0.13}_{-0.08}$ & $0.21^{+0.13}_{-0.08}$ & $0.08^{+0.08}_{-0.04}$ \\
$1 \lesssim z \leq 2.5$ & $8.5<\text{log}(M_*/M{\odot})<10.9$ & $0.44^{+0.09}_{-0.09}$ & $0.15^{+0.09}_{-0.06}$ & $0.41^{+0.09}_{-0.09}$ \\
$2.5 < z \leq 4$ & $7<\text{log}(M_*/M{\odot})<8.5$ & $0.61^{+0.09}_{-0.09}$ & $0.27^{+0.09}_{-0.09}$ & $0.12^{+0.06}_{-0.06}$ \\
$2.5 < z \leq 4$ & $8.5<\text{log}(M_*/M{\odot})<10.9$ & $0.32^{+0.05}_{-0.06}$ & $0.28^{+0.06}_{-0.05}$ & $0.41^{+0.06}_{-0.06}$ \\
$4 < z \leq 5.5$ & $7<\text{log}(M_*/M{\odot})<8.5$ & $0.41^{+0.08}_{-0.06}$ & $0.33^{+0.08}_{-0.08}$ & $0.27^{+0.06}_{-0.06}$ \\
$4 < z \leq 5.5$ & $8.5<\text{log}(M_*/M{\odot})<10.9$ & $0.30^{+0.07}_{-0.06}$ & $0.30^{+0.07}_{-0.07}$ & $0.41^{+0.06}_{-0.07}$ \\
$5.5 < z < 7$ & $7<\text{log}(M_*/M{\odot})<8.5$ & $0.53^{+0.06}_{-0.09}$ & $0.28^{+0.09}_{-0.06}$ & $0.19^{+0.06}_{-0.06}$ \\
$5.5 < z < 7$ & $8.5<\text{log}(M_*/M{\odot})<10.9$ & $0.26^{+0.13}_{-0.09}$ & $0.30^{+0.13}_{-0.13}$ & $0.43^{+0.13}_{-0.09}$ \\
\enddata
\end{deluxetable*}

\subsection{The Duty Cycle of Star Formation} \label{sec:duty_cycle}

We also consider how the duty cycle of star formation---probed through these fractional values---evolves with redshift. \cite{Clarke2024} found their data reveal no consistent trend with redshift, but suggests a low duty cycle ($f_{\text{below}}>f_{\text{above}}$) between $z=1.4$--$4$ and a high duty cycle ($f_{\text{above}}>f_{\text{below}}$) between $z=4$--$5$. As seen in Figure~\ref{fig:lum_ratio_fraction_around_equilibrium}, our results provide new insight into the galaxy populations driving these trends. We find that higher-mass galaxies maintain a low duty cycle across all redshifts probed except $1 \leq z < 2.5$, while lower-mass galaxies exhibit a high duty cycle at all redshifts. Our results are broadly consistent with recent observational work using CEERS data. \citet{Cole2025} examined SFHs for galaxies at $4.5 < z \leq 9$ and found that, on average, lower-mass galaxies ($8.7\leq\text{log}(M_*/M_{\odot})<9.3$) exhibit SFHs that are approximately constant over 10--100~Myr timescales, whereas higher-mass galaxies ($9.3\leq\text{log}(M_*/M_{\odot})\leq10.3$) tend to show declining SFHs over the same periods. In their highest-redshift bins ($6 < z \leq 7$ and $7< z \leq 9$), they report that the fraction of lower-mass galaxies with rising recent SFRs ($\log(\text{SFR}_{10}/\text{SFR}_{100}) > 0$) exceeds 50\%, consistent with burst-dominated SFHs in this regime.

Furthermore, we compute the redshift-binned difference in the fractional values between $z=1$--$2.5$ and $z=5.5$--$7$, as summarized in Table~\ref{tab:fractional_change}. These results indicate that there is no evolution in the duty cycle across $1\lesssim z<7$. These findings suggest that the relative balance between bursty and steady star formation remains roughly constant over $z=1$--$7$.

\begin{deluxetable}{cccc}[ht!]
\tablecaption{Fractional Shift in Galaxies Around the H$\alpha$-to-UV Ratio Equilibrium Range Between $1 \lesssim z \leq 2.5$ and $5.5 < z < 7$}
\tablehead{Stellar Mass & $\Delta f_{\text{above}}$ & $\Delta f_{\text{eq}}$ & $\Delta f_{\text{below}}$}

\startdata
$7<\text{log}(M_*/M{\odot})<8.5$ & $-0.13^{+0.15}_{-0.10}$ & $0.07^{+0.14}_{-0.12}$ & $0.11^{+0.10}_{-0.08}$ \\
$8.5\leq\text{log}(M_*/M{\odot})<10.9$ & $-0.18^{+0.12}_{-0.16}$ & $0.15^{+0.16}_{-0.14}$ & $0.02^{+0.16}_{-0.12}$ \\
\label{tab:fractional_change}
\enddata
\tablecomments{We define $\Delta f$ as the difference in the fraction of galaxies above, below, and within the H$\alpha$-to-UV equilibrium range between the $z=5.5$--$7$ and $z=1$--$2.5$ bins. Quoted uncertainties are propagated from the measurement errors in each bin.}
\end{deluxetable}

\subsection{Limitations and Sources of Bias} \label{sec:limitations_bias}

To contextualize these findings, we next consider how selection effects may influence our measurements, particularly for low-mass galaxies at high redshift. Galaxies experiencing a burst of star formation are bright and easy to detect, whereas galaxies in quiescent or post-burst phases are difficult to observe, especially at high redshift. Even after a recent burst, the most massive stars in a galaxy die off within $\sim$10~Myr, causing both the nebular emission lines and the UV continuum to fade rapidly. This effect is more pronounced for low- to moderate-mass galaxies at high redshift. Using the high-redshift suite \citep{Ma2018a, Ma2018b} of the second-generation Feedback in Realistic Environments (FIRE-2; \citealt{Hopkins2018}) cosmological zoom-in simulations, \cite{Sun2023a} estimated only 50\% of moderate mass ($M_*/M_{\odot} \sim 10^{8.5}-10^9$) galaxies at $z\sim7$ can be detected by surveys such as JADES and CEERS and that galaxies near this detection limit tend to have experienced a recent burst of star formation. As a result, our lower-mass ($7\leq\text{log}(M_*/M_{\odot})<8.5$), high-redshift ($z = 5.5$--$7$) sample is likely biased toward galaxies experiencing bursts within the past $\sim$10~Myr or galaxies undergoing frequent, short-timescale bursts every $\sim 100$~Myr \citep{Mehta2023}. For CEERS galaxies, this bias is exacerbated by the NIRSpec slit selection, which for $z<7$ galaxies predominantly targets UV-bright sources. Together, these selection effects may artificially inflate the observed burstiness of low-mass, high-redshift galaxies by underrepresenting quiescent or post-burst systems. Consequently, the true fraction of galaxies with steady or recently declining SFHs is likely higher than observed, and our measured values of $f_{\rm above}$ and $\sigma_{\rm int}$ may be skewed toward more extreme, burst-dominated populations. These trends should be interpreted with caution, as it may lead to overestimating the presence of bursty star formation in the early Universe. 

Furthermore, while our analysis removes broad-line AGN, we do not explicitly exclude narrow-line (NL) AGN from the sample. NL AGN, which may not significantly contribute to the rest-frame UV emission due to the obscuration and orientation of these systems, could nonetheless enhance nebular line emission such as H$\alpha$, potentially mimicking elevated SFRs. Although the impact is likely limited, especially given the rarity of luminous AGN at these masses and redshifts, the potential contamination by NL AGN should be considered when interpreting the results of this work.

Finally, we note that several key modeling assumptions may introduce systematic uncertainties that could influence our results. These include (1) adopting a single attenuation law (i.e., \citet{Calzetti2001}) for all galaxies, (2) assuming Case B recombination with fixed nebular conditions, and (3) applying a constant ratio between nebular and continuum attenuation across the full redshift and stellar mass range. While these assumptions are motivated by consistency with previous literature, the magnitude of their potential impact may be comparable to the differences we observe in H$\alpha$-to-UV ratios and SFMS scatter across redshift and stellar mass. In particular, recent works indicate that the attenuation-curve slope evolves with redshift, becoming steeper at $z \lesssim 3$ and flatter at $z \gtrsim 5$ \citep[e.g.,][]{Markov2025, Shivaei2025}. While we do not observe a redshift evolution over $z=1$--$7$ in galaxy burstiness (as probed by the H$\alpha$-to-UV ratio), adopting a redshift-dependent attenuation law (i.e., Equations 5 and 6 from \citet{Shivaei2025}) could introduce one. A steeper curve at low redshift would lower the H$\alpha$-to-UV ratio, suppressing inferred burstiness, whereas a flatter curve at high redshift would raise the H$\alpha$-to-UV ratio, increasing inferred burstiness. Future work should aim to quantify the sensitivity of burstiness indicators to these assumptions.

\section{Summary} \label{sec:summary}

In this work, we analyze the SFMS and the H$\alpha$-to-UV ratio of a statistically robust sample of 346 star-forming galaxies at $1 \lesssim z < 7$ in the CEERS and RUBIES spectroscopic surveys to investigate the burstiness of star formation at these redshifts. Our main conclusions are as follows:
\begin{enumerate}
    \item We find that the difference in intrinsic scatter between the H$\alpha$-based SFMS and UV-continuum-based SFMS ${\langle \sigma_{\rm{}H\alpha} \rangle} - {\langle \sigma_{\rm{}UV} \rangle} = 0.01 \pm 0.04$~dex, is consistent with zero and therefore not statistically significant for our stellar-mass complete subsample.
    \item The mass-incomplete subsample exhibits visually flatter H$\alpha$-based SFMS relations compared to the best-fit relations given by the mass-complete subsample and literature relations, especially at high redshifts ($z \gtrsim 5.5$). This may be due to a combination of (1) NIRSpec sensitivity limits that bias against low-SFR galaxies---artificially raising the average observed SFR---and (2) elevated H$\alpha$ emission from galaxies undergoing bursts of star formation. These effects limit the utility of the SFMS in identifying individual bursty galaxies and motivate a complementary diagnostic approach. 
    \item To more directly probe recent ($\lesssim100$~Myr) SFR variability, we examine the dust-corrected H$\alpha$-to-UV ratio. We find evidence for bursty star formation in $73^{+4}_{-4}$~\% of galaxies in our sample based on their H$\alpha$-to-UV ratios, calculated as the combined fraction of sources above and below the equilibrium range defined by \citet{Mehta2023}. We find no variation in this total bursty fraction between our lower-mass and higher-mass subsamples. However, the distribution of bursty systems differs with mass: a higher fraction of lower-mass galaxies lie above the equilibrium range ($f_{\text{above}}= 0.53^{+0.05}_{-0.05}$) and a lower fraction below it ($f_{\text{below}} = 0.18^{+0.04}_{-0.03}$) compared to higher-mass galaxies with $f_{\text{above}}= 0.33^{+0.04}_{-0.04}$ and $f_{\text{below}} = 0.41^{+0.04}_{-0.04}$. Thus, we find lower-mass galaxies are $30 \pm 1$\% more likely to be observed in a recent ($\lesssim 100$~Myr) burst of star formation compared to their higher-mass counterparts.
    \item We infer the evolutionary pathway of galaxies around the H$\alpha$-to-UV equilibrium range by computing the fractional change $\Delta f$ in H$\alpha$-to-UV ratios between $z=1$--$2.5$ and $z=5.5$--$7$. Across both mass bins, we find that $\Delta f$ values are statistically consistent with zero, suggesting little to no evolution in the relative fraction of galaxies above, below, or within the equilibrium range over $z=1$--$7$. 
\end{enumerate}
With the upcoming \textit{JWST} surveys such as the CAPERS program (\textit{JWST} Cycle 3 GO\#6368, PI Dickinson; \citealt{Dickinson2024}), future robust analyses breaking upward the redshift range of the current paper are possible with \textit{JWST} and would probe the SFH of the earliest galaxies in the Universe.

\begin{acknowledgments}

The authors would like to thank Bahram Mobasher, Benne Holwerda, Mauro Giavalisco, Michaela Hirschmann, and Stephen Wilkins for thorough and detailed feedback, which helped improve the quality of this manuscript. We thank the RUBIES team (including PI's Anna de Graaff and Gabriel Brammer) for carrying out this program and making the data publicly available. MNP acknowledges support for the Galaxy Evolution Vertically Integrated Project (GEVIP) undergraduate research group from the National Science Foundation under NSF-AAG-1908817 and NSF-AAG-2009905. 

The \textit{JWST} data presented in this article were obtained from the Mikulski Archive for Space Telescopes (MAST) at the Space Telescope Science Institute. The specifc observations
analyzed can be accessed via doi:\href{https://doi.org/10.17909/0eta-ym85}{10.17909/0eta-ym85}.

\end{acknowledgments}

\newpage
\bibliography{main}{}

\begin{thebibliography}{}
\expandafter\ifx\csname natexlab\endcsname\relax\def\natexlab#1{#1}\fi
\providecommand{\url}[1]{\href{#1}{#1}}
\providecommand{\dodoi}[1]{doi:~\href{http://doi.org/#1}{\nolinkurl{#1}}}
\providecommand{\doeprint}[1]{\href{http://ascl.net/#1}{\nolinkurl{http://ascl.net/#1}}}
\providecommand{\doarXiv}[1]{\href{https://arxiv.org/abs/#1}{\nolinkurl{https://arxiv.org/abs/#1}}}

\bibitem[{P. {Arrabal Haro} {et~al.}(2023{\natexlab{a}}){Arrabal Haro}, {Dickinson}, {Finkelstein}, {Fujimoto}, {Fern{\'a}ndez}, {Kartaltepe}, {Jung}, {Cole}, {Burgarella}, {Chworowsky}, {Hutchison}, {Morales}, {Papovich}, {Simons}, {Amor{\'\i}n}, {Backhaus}, {Bagley}, {Bisigello}, {Calabr{\`o}}, {Castellano}, {Cleri}, {Dav{\'e}}, {Dekel}, {Ferguson}, {Fontana}, {Gawiser}, {Giavalisco}, {Harish}, {Hathi}, {Hirschmann}, {Holwerda}, {Huertas-Company}, {Koekemoer}, {Larson}, {Lucas}, {Mobasher}, {P{\'e}rez-Gonz{\'a}lez}, {Pirzkal}, {Rose}, {Santini}, {Trump}, {de la Vega}, {Wang}, {Weiner}, {Wilkins}, {Yang}, {Yung}, \& {Zavala}}]{ArrabelHaro2023ApJ}
{Arrabal Haro}, P., {Dickinson}, M., {Finkelstein}, S.~L., {et~al.} 2023{\natexlab{a}}, \bibinfo{title}{{Spectroscopic Confirmation of CEERS NIRCam-selected Galaxies at z ≃ 8-10},} \apjl, 951, L22, \dodoi{10.3847/2041-8213/acdd54}

\bibitem[{P. {Arrabal Haro} {et~al.}(2023{\natexlab{b}}){Arrabal Haro}, {Dickinson}, {Finkelstein}, {Kartaltepe}, {Donnan}, {Burgarella}, {Carnall}, {Cullen}, {Dunlop}, {Fern{\'a}ndez}, {Fujimoto}, {Jung}, {Krips}, {Larson}, {Papovich}, {P{\'e}rez-Gonz{\'a}lez}, {Amor{\'\i}n}, {Bagley}, {Buat}, {Casey}, {Chworowsky}, {Cohen}, {Ferguson}, {Giavalisco}, {Huertas-Company}, {Hutchison}, {Kocevski}, {Koekemoer}, {Lucas}, {McLeod}, {McLure}, {Pirzkal}, {Seill{\'e}}, {Trump}, {Weiner}, {Wilkins}, \& {Zavala}}]{ArrabelHaro2023Nat}
{Arrabal Haro}, P., {Dickinson}, M., {Finkelstein}, S.~L., {et~al.} 2023{\natexlab{b}}, \bibinfo{title}{{Confirmation and refutation of very luminous galaxies in the early Universe},} \nat, 622, 707, \dodoi{10.1038/s41586-023-06521-7}

\bibitem[{Y. {Asada} {et~al.}(2024){Asada}, {Sawicki}, {Abraham}, {Brada{\v{c}}}, {Brammer}, {Desprez}, {Estrada-Carpenter}, {Iyer}, {Martis}, {Matharu}, {Mowla}, {Muzzin}, {Noirot}, {Sarrouh}, {Strait}, {Willott}, \& {Harshan}}]{Asada2024}
{Asada}, Y., {Sawicki}, M., {Abraham}, R., {et~al.} 2024, \bibinfo{title}{{Bursty star formation and galaxy-galaxy interactions in low-mass galaxies 1 Gyr after the Big Bang},} \mnras, 527, 11372, \dodoi{10.1093/mnras/stad3902}

\bibitem[{R. {Asquith} {et~al.}(2018){Asquith}, {Pearce}, {Almaini}, {Knebe}, {Gonzalez-Perez}, {Benson}, {Blaizot}, {Carretero}, {Castander}, {Cattaneo}, {Cora}, {Croton}, {Devriendt}, {Fontanot}, {Gargiulo}, {Hartley}, {Henriques}, {Lee}, {Mamon}, {Onions}, {Padilla}, {Power}, {Srisawat}, {Stevens}, {Thomas}, {Vega-Mart{\'\i}nez}, \& {Yi}}]{Asquith2018}
{Asquith}, R., {Pearce}, F.~R., {Almaini}, O., {et~al.} 2018, \bibinfo{title}{{Cosmic CARNage II: the evolution of the galaxy stellar mass function in observations and galaxy formation models},} \mnras, 480, 1197, \dodoi{10.1093/mnras/sty1870}

\bibitem[{H. {Atek} {et~al.}(2022){Atek}, {Furtak}, {Oesch}, {van Dokkum}, {Reddy}, {Contini}, {Illingworth}, \& {Wilkins}}]{Atek2022}
{Atek}, H., {Furtak}, L.~J., {Oesch}, P., {et~al.} 2022, \bibinfo{title}{{The star formation burstiness and ionizing efficiency of low-mass galaxies},} \mnras, 511, 4464, \dodoi{10.1093/mnras/stac360}

\bibitem[{M.~B. {Bagley} {et~al.}(2023){Bagley}, {Finkelstein}, {Koekemoer}, {Ferguson}, {Arrabal Haro}, {Dickinson}, {Kartaltepe}, {Papovich}, {P{\'e}rez-Gonz{\'a}lez}, {Pirzkal}, {Somerville}, {Willmer}, {Yang}, {Yung}, {Fontana}, {Grazian}, {Grogin}, {Hirschmann}, {Kewley}, {Kirkpatrick}, {Kocevski}, {Lotz}, {Medrano}, {Morales}, {Pentericci}, {Ravindranath}, {Trump}, {Wilkins}, {Calabr{\`o}}, {Cooper}, {Costantin}, {de la Vega}, {Hilbert}, {Hutchison}, {Larson}, {Lucas}, {McGrath}, {Ryan}, {Wang}, \& {Wuyts}}]{Bagley2023}
{Bagley}, M.~B., {Finkelstein}, S.~L., {Koekemoer}, A.~M., {et~al.} 2023, \bibinfo{title}{{CEERS Epoch 1 NIRCam Imaging: Reduction Methods and Simulations Enabling Early JWST Science Results},} \apjl, 946, L12, \dodoi{10.3847/2041-8213/acbb08}

\bibitem[{A. {Broussard} {et~al.}(2022){Broussard}, {Gawiser}, \& {Iyer}}]{Broussard2022}
{Broussard}, A., {Gawiser}, E., \& {Iyer}, K. 2022, \bibinfo{title}{{Improved Measurements of Galaxy Star Formation Stochasticity from the Intrinsic Scatter of Burst Indicators},} \apj, 939, 35, \dodoi{10.3847/1538-4357/ac94c2}

\bibitem[{A. {Broussard} {et~al.}(2019){Broussard}, {Gawiser}, {Iyer}, {Kurczynski}, {Somerville}, {Dav{\'e}}, {Finkelstein}, {Jung}, \& {Pacifici}}]{Broussard2019}
{Broussard}, A., {Gawiser}, E., {Iyer}, K., {et~al.} 2019, \bibinfo{title}{{Star Formation Stochasticity Measured from the Distribution of Burst Indicators},} \apj, 873, 74, \dodoi{10.3847/1538-4357/ab04ad}

\bibitem[{H. {Bushouse} {et~al.}(2022){Bushouse}, {Eisenhamer}, {Dencheva}, {Davies}, {Greenfield}, {Morrison}, {Hodge}, {Simon}, {Grumm}, {Droettboom}, {Slavich}, {Sosey}, {Pauly}, {Miller}, {Jedrzejewski}, {Hack}, {Davis}, {Crawford}, {Law}, {Gordon}, {Regan}, {Cara}, {MacDonald}, {Bradley}, {Shanahan}, {Jamieson}, {Teodoro}, \& {Williams}}]{Bushouse2022_JWST_Pipeline}
{Bushouse}, H., {Eisenhamer}, J., {Dencheva}, N., {et~al.} 2022, \bibinfo{title}{{JWST Calibration Pipeline},}, 1.8.5 Zenodo, \dodoi{10.5281/zenodo.7429939}

\bibitem[{D. {Calzetti}(2001){Calzetti}}]{Calzetti2001}
{Calzetti}, D. 2001, \bibinfo{title}{{The Dust Opacity of Star-forming Galaxies},} \pasp, 113, 1449, \dodoi{10.1086/324269}

\bibitem[{D. {Calzetti} {et~al.}(2000){Calzetti}, {Armus}, {Bohlin}, {Kinney}, {Koornneef}, \& {Storchi-Bergmann}}]{Calzetti2000}
{Calzetti}, D., {Armus}, L., {Bohlin}, R.~C., {et~al.} 2000, \bibinfo{title}{{The Dust Content and Opacity of Actively Star-forming Galaxies},} \apj, 533, 682, \dodoi{10.1086/308692}

\bibitem[{A.~C. {Carnall} {et~al.}(2018){Carnall}, {McLure}, {Dunlop}, \& {Dav{\'e}}}]{carnall18}
{Carnall}, A.~C., {McLure}, R.~J., {Dunlop}, J.~S., \& {Dav{\'e}}, R. 2018, \bibinfo{title}{{Inferring the star formation histories of massive quiescent galaxies with BAGPIPES: evidence for multiple quenching mechanisms},} \mnras, 480, 4379, \dodoi{10.1093/mnras/sty2169}

\bibitem[{A.~C. {Carnall} {et~al.}(2019){Carnall}, {McLure}, {Dunlop}, {Cullen}, {McLeod}, {Wild}, {Johnson}, {Appleby}, {Dav{\'e}}, {Amorin}, {Bolzonella}, {Castellano}, {Cimatti}, {Cucciati}, {Gargiulo}, {Garilli}, {Marchi}, {Pentericci}, {Pozzetti}, {Schreiber}, {Talia}, \& {Zamorani}}]{carnall19}
{Carnall}, A.~C., {McLure}, R.~J., {Dunlop}, J.~S., {et~al.} 2019, \bibinfo{title}{{The VANDELS survey: the star-formation histories of massive quiescent galaxies at 1.0 < z < 1.3},} \mnras, 490, 417, \dodoi{10.1093/mnras/stz2544}

\bibitem[{C. {Carvajal-Bohorquez} {et~al.}(2025){Carvajal-Bohorquez}, {Ciesla}, {Laporte}, {Boquien}, {Buat}, {Ilbert}, {Aufort}, {Shuntov}, {Witten}, {Oesch}, \& {Covelo-Paz}}]{Carvajal-Bohorquez2025}
{Carvajal-Bohorquez}, C., {Ciesla}, L., {Laporte}, N., {et~al.} 2025, \bibinfo{title}{{Stochastic star formation activity of galaxies within the first billion years probed by JWST},} arXiv e-prints, arXiv:2507.13160, \dodoi{10.48550/arXiv.2507.13160}

\bibitem[{G. Chabrier(2003)Chabrier}]{Chabrier2003}
Chabrier, G. 2003, \bibinfo{title}{Galactic Stellar and Substellar Initial Mass Function1,} Publications of the Astronomical Society of the Pacific, 115, 763, \dodoi{10.1086/376392}

\bibitem[{K. {Chworowsky} {et~al.}(2024){Chworowsky}, {Finkelstein}, {Boylan-Kolchin}, {McGrath}, {Iyer}, {Papovich}, {Dickinson}, {Taylor}, {Yung}, {Arrabal Haro}, {Bagley}, {Backhaus}, {Bhatawdekar}, {Cheng}, {Cleri}, {Cole}, {Cooper}, {Costantin}, {Dekel}, {Franco}, {Fujimoto}, {Hayward}, {Holwerda}, {Huertas-Company}, {Hirschmann}, {Hutchison}, {Koekemoer}, {Larson}, {Li}, {Long}, {Lucas}, {Pirzkal}, {Rodighiero}, {Somerville}, {Vanderhoof}, {de la Vega}, {Wilkins}, {Yang}, \& {Zavala}}]{Chworowsky2024}
{Chworowsky}, K., {Finkelstein}, S.~L., {Boylan-Kolchin}, M., {et~al.} 2024, \bibinfo{title}{{Evidence for a Shallow Evolution in the Volume Densities of Massive Galaxies at z = 4{\textendash}8 from CEERS},} \aj, 168, 113, \dodoi{10.3847/1538-3881/ad57c1}

\bibitem[{L. {Ciesla} {et~al.}(2024){Ciesla}, {Elbaz}, {Ilbert}, {Buat}, {Magnelli}, {Narayanan}, {Daddi}, {G{\'o}mez-Guijarro}, \& {Arango-Toro}}]{Ciesla2024}
{Ciesla}, L., {Elbaz}, D., {Ilbert}, O., {et~al.} 2024, \bibinfo{title}{{Identification of a transition from stochastic to secular star formation around z = 9 with JWST},} \aap, 686, A128, \dodoi{10.1051/0004-6361/202348091}

\bibitem[{L. {Clarke} {et~al.}(2024){Clarke}, {Shapley}, {Sanders}, {Topping}, {Brammer}, {Bento}, {Reddy}, \& {Kehoe}}]{Clarke2024}
{Clarke}, L., {Shapley}, A.~E., {Sanders}, R.~L., {et~al.} 2024, \bibinfo{title}{{The Star-forming Main Sequence in JADES and CEERS at z > 1.4: Investigating the Burstiness of Star Formation},} \apj, 977, 133, \dodoi{10.3847/1538-4357/ad8ba4}

\bibitem[{N.~J. {Cleri} {et~al.}(2022){Cleri}, {Trump}, {Backhaus}, {Momcheva}, {Papovich}, {Simons}, {Weiner}, {Estrada-Carpenter}, {Finkelstein}, {Giavalisco}, {Ji}, {Jung}, {Matharu}, {Martinez}, \& {Sturm}}]{Cleri2022}
{Cleri}, N.~J., {Trump}, J.~R., {Backhaus}, B.~E., {et~al.} 2022, \bibinfo{title}{{CLEAR: Paschen-{\ensuremath{\beta}} Star Formation Rates and Dust Attenuation of Low-redshift Galaxies},} \apj, 929, 3, \dodoi{10.3847/1538-4357/ac5a4c}

\bibitem[{J.~W. {Cole} {et~al.}(2025){Cole}, {Papovich}, {Finkelstein}, {Bagley}, {Dickinson}, {Iyer}, {Yung}, {Ciesla}, {Amor{\'\i}n}, {Arrabal Haro}, {Bhatawdekar}, {Calabr{\`o}}, {Cleri}, {de la Vega}, {Dekel}, {Endsley}, {Gawiser}, {Giavalisco}, {Hathi}, {Hirschmann}, {Holwerda}, {Kartaltepe}, {Koekemoer}, {Lucas}, {Mascia}, {Mobasher}, {P{\'e}rez-Gonz{\'a}lez}, {Rodighiero}, {Ronayne}, {Tacchella}, {Weiner}, \& {Wilkins}}]{Cole2025}
{Cole}, J.~W., {Papovich}, C., {Finkelstein}, S.~L., {et~al.} 2025, \bibinfo{title}{{CEERS: Increasing Scatter along the Star-forming Main Sequence Indicates Early Galaxies Form in Bursts},} \apj, 979, 193, \dodoi{10.3847/1538-4357/ad9a6a}

\bibitem[{C. {Conselice} {et~al.}(2023){Conselice}, {Adams}, {Austin}, {Trussler}, \& {De Albernaz Ferreira}}]{Conselice2023}
{Conselice}, C., {Adams}, N., {Austin}, D., {Trussler}, J., \& {De Albernaz Ferreira}, L. 2023, in American Astronomical Society Meeting Abstracts, Vol. 241, American Astronomical Society Meeting Abstracts, 153.07

\bibitem[{E. {Daddi} {et~al.}(2007){Daddi}, {Dickinson}, {Morrison}, {Chary}, {Cimatti}, {Elbaz}, {Frayer}, {Renzini}, {Pope}, {Alexander}, {Bauer}, {Giavalisco}, {Huynh}, {Kurk}, \& {Mignoli}}]{Daddi2007}
{Daddi}, E., {Dickinson}, M., {Morrison}, G., {et~al.} 2007, \bibinfo{title}{{Multiwavelength Study of Massive Galaxies at z\raisebox{-0.5ex}\textasciitilde2. I. Star Formation and Galaxy Growth},} \apj, 670, 156, \dodoi{10.1086/521818}

\bibitem[{K. {Daikuhara} {et~al.}(2024){Daikuhara}, {Kodama}, {P{\'e}rez-Mart{\'\i}nez}, {Shimakawa}, {Suzuki}, {Tadaki}, {Koyama}, \& {Tanaka}}]{Daikuhara2024}
{Daikuhara}, K., {Kodama}, T., {P{\'e}rez-Mart{\'\i}nez}, J.~M., {et~al.} 2024, \bibinfo{title}{{Star-formation activity of low-mass galaxies at the peak epoch of galaxy formation probed by deep narrow-band imaging},} \mnras, 531, 2335, \dodoi{10.1093/mnras/stae1243}

\bibitem[{P. {Dayal} {et~al.}(2013){Dayal}, {Dunlop}, {Maio}, \& {Ciardi}}]{Dayal2013}
{Dayal}, P., {Dunlop}, J.~S., {Maio}, U., \& {Ciardi}, B. 2013, \bibinfo{title}{{Simulating the assembly of galaxies at redshifts z = 6-12},} \mnras, 434, 1486, \dodoi{10.1093/mnras/stt1108}

\bibitem[{A. {de Graaff} {et~al.}(2025){de Graaff}, {Brammer}, {Weibel}, {Lewis}, {Maseda}, {Oesch}, {Bezanson}, {Boogaard}, {Cleri}, {Cooper}, {Gottumukkala}, {Greene}, {Hirschmann}, {Hviding}, {Katz}, {Labb{\'e}}, {Leja}, {Matthee}, {McConachie}, {Miller}, {Naidu}, {Price}, {Rix}, {Setton}, {Suess}, {Wang}, {Whitaker}, \& {Williams}}]{deGraaff2025}
{de Graaff}, A., {Brammer}, G., {Weibel}, A., {et~al.} 2025, \bibinfo{title}{{RUBIES: A complete census of the bright and red distant Universe with JWST/NIRSpec},} \aap, 697, A189, \dodoi{10.1051/0004-6361/202452186}

\bibitem[{A.~G. {de Graaff} {et~al.}(2023){de Graaff}, {Brammer}, {Bezanson}, {Gould}, {Hirschmann}, {Katz}, {Labbe}, {Leja}, {Lewis}, {Maseda}, {Matthee}, {Naidu}, {Oesch}, {Rix}, {Setton}, {Shivaei}, {Suess}, {Whitaker}, \& {Williams}}]{Graaff2023}
{de Graaff}, A.~G., {Brammer}, G., {Bezanson}, R., {et~al.} 2023, \bibinfo{title}{{A complete census of the rare, extreme and red: a NIRCam-selected extragalactic community survey with JWST/NIRSpec},}, JWST Proposal. Cycle 2, ID. \#4233

\bibitem[{M. {Dickinson} {et~al.}(2024){Dickinson}, {Amorin}, {Arrabal Haro}, {Bagley}, {Barro}, {Buat}, {Burgarella}, {Calabro'}, {Carnall}, {Casey}, {Chworowsky}, {Cleri}, {Cole}, {Cooper}, {Cullen}, {Daddi}, {Donnan}, {Dunlop}, {Elbaz}, {Ferguson}, {Fernandez}, {Finkelstein}, {Fontana}, {Fujimoto}, {Giavalisco}, {Hamilton}, {Hathi}, {Hirschmann}, {Hutchison}, {Juneau}, {Jung}, {Kartaltepe}, {Kocevski}, {Koekemoer}, {Larson}, {Long}, {Lucas}, {Mascia}, {McGrath}, {McLeod}, {McLure}, {Napolitano}, {Papovich}, {Pentericci}, {Perez Gonzalez}, {Simons}, {Somerville}, {Trump}, {Wang}, {Weiner}, {Wilkins}, {Yung}, \& {Zavala}}]{Dickinson2024}
{Dickinson}, M., {Amorin}, R., {Arrabal Haro}, P., {et~al.} 2024, \bibinfo{title}{{The CANDELS-Area Prism Epoch of Reionization Survey (CAPERS)},}, JWST Proposal. Cycle 3, ID. \#6368

\bibitem[{C.~T. {Donnan} {et~al.}(2024){Donnan}, {McLure}, {Dunlop}, {McLeod}, {Magee}, {Arellano-C{\'o}rdova}, {Barrufet}, {Begley}, {Bowler}, {Carnall}, {Cullen}, {Ellis}, {Fontana}, {Illingworth}, {Grogin}, {Hamadouche}, {Koekemoer}, {Liu}, {Mason}, {Santini}, \& {Stanton}}]{Donnan2024}
{Donnan}, C.~T., {McLure}, R.~J., {Dunlop}, J.~S., {et~al.} 2024, \bibinfo{title}{{JWST PRIMER: a new multifield determination of the evolving galaxy UV luminosity function at redshifts z ≃ 9 - 15},} \mnras, 533, 3222, \dodoi{10.1093/mnras/stae2037}

\bibitem[{D. {Elbaz} {et~al.}(2007){Elbaz}, {Daddi}, {Le Borgne}, {Dickinson}, {Alexander}, {Chary}, {Starck}, {Brandt}, {Kitzbichler}, {MacDonald}, {Nonino}, {Popesso}, {Stern}, \& {Vanzella}}]{Elbaz2007}
{Elbaz}, D., {Daddi}, E., {Le Borgne}, D., {et~al.} 2007, \bibinfo{title}{{The reversal of the star formation-density relation in the distant universe},} \aap, 468, 33, \dodoi{10.1051/0004-6361:20077525}

\bibitem[{N. {Emami} {et~al.}(2019){Emami}, {Siana}, {Weisz}, {Johnson}, {Ma}, \& {El-Badry}}]{Emami2019}
{Emami}, N., {Siana}, B., {Weisz}, D.~R., {et~al.} 2019, \bibinfo{title}{{A Closer Look at Bursty Star Formation with L $_{H{\ensuremath{\alpha}} }$ and L $_{UV}$ Distributions},} \apj, 881, 71, \dodoi{10.3847/1538-4357/ab211a}

\bibitem[{R. {Endsley} {et~al.}(2024){Endsley}, {Chisholm}, {Stark}, {Topping}, \& {Whitler}}]{Endsley2024}
{Endsley}, R., {Chisholm}, J., {Stark}, D.~P., {Topping}, M.~W., \& {Whitler}, L. 2024, \bibinfo{title}{{The Burstiness of Star Formation at $z\sim6$: A Huge Diversity in the Recent Star Formation Histories of Very UV-faint Galaxies},} arXiv e-prints, arXiv:2410.01905, \dodoi{10.48550/arXiv.2410.01905}

\bibitem[{A.~L. {Faisst} {et~al.}(2019){Faisst}, {Capak}, {Emami}, {Tacchella}, \& {Larson}}]{Faisst2019}
{Faisst}, A.~L., {Capak}, P.~L., {Emami}, N., {Tacchella}, S., \& {Larson}, K.~L. 2019, \bibinfo{title}{{The Recent Burstiness of Star Formation in Galaxies at z {\ensuremath{\sim}} 4.5 from H{\ensuremath{\alpha}} Measurements},} \apj, 884, 133, \dodoi{10.3847/1538-4357/ab425b}

\bibitem[{V. {Fern{\'a}ndez} {et~al.}(2024){Fern{\'a}ndez}, {Amor{\'\i}n}, {Firpo}, \& {Morisset}}]{Fernandez2024}
{Fern{\'a}ndez}, V., {Amor{\'\i}n}, R., {Firpo}, V., \& {Morisset}, C. 2024, \bibinfo{title}{{LIME: A LIne MEasuring library for large and complex spectroscopic data sets. I. Implementation of a virtual observatory for JWST spectra},} \aap, 688, A69, \dodoi{10.1051/0004-6361/202449224}

\bibitem[{S.~L. {Finkelstein} {et~al.}(2017){Finkelstein}, {Dickinson}, {Ferguson}, {Grazian}, {Grogin}, {Kartaltepe}, {Kewley}, {Kocevski}, {Koekemoer}, {Lotz}, {Papovich}, {Pentericci}, {Perez-Gonzalez}, {Pirzkal}, {Ravindranath}, {Somerville}, {Trump}, \& {Wilkins}}]{Finkelstein2017}
{Finkelstein}, S.~L., {Dickinson}, M., {Ferguson}, H.~C., {et~al.} 2017, \bibinfo{title}{{The Cosmic Evolution Early Release Science (CEERS) Survey},}, JWST Proposal ID 1345. Cycle 0 Early Release Science

\bibitem[{S.~L. {Finkelstein} {et~al.}(2023){Finkelstein}, {Bagley}, {Ferguson}, {Wilkins}, {Kartaltepe}, {Papovich}, {Yung}, {Arrabal Haro}, {Behroozi}, {Dickinson}, {Kocevski}, {Koekemoer}, {Larson}, {Le Bail}, {Morales}, {P{\'e}rez-Gonz{\'a}lez}, {Burgarella}, {Dav{\'e}}, {Hirschmann}, {Somerville}, {Wuyts}, {Bromm}, {Casey}, {Fontana}, {Fujimoto}, {Gardner}, {Giavalisco}, {Grazian}, {Grogin}, {Hathi}, {Hutchison}, {Jha}, {Jogee}, {Kewley}, {Kirkpatrick}, {Long}, {Lotz}, {Pentericci}, {Pierel}, {Pirzkal}, {Ravindranath}, {Ryan}, {Trump}, {Yang}, {Bhatawdekar}, {Bisigello}, {Buat}, {Calabr{\`o}}, {Castellano}, {Cleri}, {Cooper}, {Croton}, {Daddi}, {Dekel}, {Elbaz}, {Franco}, {Gawiser}, {Holwerda}, {Huertas-Company}, {Jaskot}, {Leung}, {Lucas}, {Mobasher}, {Pandya}, {Tacchella}, {Weiner}, \& {Zavala}}]{Finkelstein2023}
{Finkelstein}, S.~L., {Bagley}, M.~B., {Ferguson}, H.~C., {et~al.} 2023, \bibinfo{title}{{CEERS Key Paper. I. An Early Look into the First 500 Myr of Galaxy Formation with JWST},} \apjl, 946, L13, \dodoi{10.3847/2041-8213/acade4}

\bibitem[{S.~L. {Finkelstein} {et~al.}(2024){Finkelstein}, {Leung}, {Bagley}, {Dickinson}, {Ferguson}, {Papovich}, {Akins}, {Arrabal Haro}, {Dav{\'e}}, {Dekel}, {Kartaltepe}, {Kocevski}, {Koekemoer}, {Pirzkal}, {Somerville}, {Yung}, {Amor{\'\i}n}, {Backhaus}, {Behroozi}, {Bisigello}, {Bromm}, {Casey}, {Ch{\'a}vez Ortiz}, {Cheng}, {Chworowsky}, {Cleri}, {Cooper}, {Davis}, {de la Vega}, {Elbaz}, {Franco}, {Fontana}, {Fujimoto}, {Giavalisco}, {Grogin}, {Holwerda}, {Huertas-Company}, {Hirschmann}, {Iyer}, {Jogee}, {Jung}, {Larson}, {Lucas}, {Mobasher}, {Morales}, {Morley}, {Mukherjee}, {P{\'e}rez-Gonz{\'a}lez}, {Ravindranath}, {Rodighiero}, {Rowland}, {Tacchella}, {Taylor}, {Trump}, \& {Wilkins}}]{Finkelstein2024}
{Finkelstein}, S.~L., {Leung}, G. C.~K., {Bagley}, M.~B., {et~al.} 2024, \bibinfo{title}{{The Complete CEERS Early Universe Galaxy Sample: A Surprisingly Slow Evolution of the Space Density of Bright Galaxies at z {\ensuremath{\sim}} 8.5{\textendash}14.5},} \apjl, 969, L2, \dodoi{10.3847/2041-8213/ad4495}

\bibitem[{S.~L. {Finkelstein} {et~al.}(2025){Finkelstein}, {Bagley}, {Arrabal Haro}, {Dickinson}, {Ferguson}, {Kartaltepe}, {Kocevski}, {Koekemoer}, {Lotz}, {Papovich}, {P{\'e}rez-Gonz{\'a}lez}, {Pirzkal}, {Somerville}, {Trump}, {Yang}, {Yung}, {Fontana}, {Grazian}, {Grogin}, {Kewley}, {Kirkpatrick}, {Larson}, {Pentericci}, {Ravindranath}, {Wilkins}, {Almaini}, {Amor{\'\i}n}, {Barro}, {Bhatawdekar}, {Bisigello}, {Brooks}, {Buat}, {Buitrago}, {Burgarella}, {Calabr{\`o}}, {Castellano}, {Cheng}, {Cleri}, {Cole}, {Cooper}, {Cooper}, {Costantin}, {Cox}, {Croton}, {Daddi}, {Davis}, {Dekel}, {Elbaz}, {Fern{\'a}ndez}, {Fujimoto}, {Gandolfi}, {Gardner}, {Gawiser}, {Giavalisco}, {G{\'o}mez-Guijarro}, {Guo}, {Gupta}, {Hathi}, {Harish}, {Henry}, {Hirschmann}, {Hu}, {Hutchison}, {Iyer}, {Jaskot}, {Jha}, {Jung}, {Kassin}, {Kokorev}, {Kurczynski}, {Leung}, {Llerena}, {Long}, {Lucas}, {Lu}, {McGrath}, {McIntosh}, {Merlin}, {Mobasher}, {Morales}, {Napolitano}, {Pacucci}, {Pandya}, {Rafelski}, {Rodighiero}, {Rose}, {Santini},
  {Seill{\'e}}, {Simons}, {Shen}, {Straughn}, {Tacchella}, {Taylor}, {Vanderhoof}, {Vega-Ferrero}, {Weiner}, {Willmer}, {Zhu}, {Bell}, {Wuyts}, {Holwerda}, {Wang}, {Wang}, {Zavala}, \& {CEERS Collaboration}}]{finkelstein2025}
{Finkelstein}, S.~L., {Bagley}, M.~B., {Arrabal Haro}, P., {et~al.} 2025, \bibinfo{title}{{The Cosmic Evolution Early Release Science Survey (CEERS)},} \apjl, 983, L4, \dodoi{10.3847/2041-8213/adbbd3}

\bibitem[{D. {Foreman-Mackey} {et~al.}(2013){Foreman-Mackey}, {Hogg}, {Lang}, \& {Goodman}}]{Foreman-Mackey2013}
{Foreman-Mackey}, D., {Hogg}, D.~W., {Lang}, D., \& {Goodman}, J. 2013, \bibinfo{title}{{emcee: The MCMC Hammer},} \pasp, 125, 306, \dodoi{10.1086/670067}

\bibitem[{J.~P. {Gardner} {et~al.}(2006){Gardner}, {Mather}, {Clampin}, {Doyon}, {Greenhouse}, {Hammel}, {Hutchings}, {Jakobsen}, {Lilly}, {Long}, {Lunine}, {McCaughrean}, {Mountain}, {Nella}, {Rieke}, {Rieke}, {Rix}, {Smith}, {Sonneborn}, {Stiavelli}, {Stockman}, {Windhorst}, \& {Wright}}]{Gardner2006}
{Gardner}, J.~P., {Mather}, J.~C., {Clampin}, M., {et~al.} 2006, \bibinfo{title}{{The James Webb Space Telescope},} \ssr, 123, 485, \dodoi{10.1007/s11214-006-8315-7}

\bibitem[{J.~P. {Gardner} {et~al.}(2023){Gardner}, {Mather}, {Abbott}, {Abell}, {Abernathy}, {Abney}, {Abraham}, {Abraham}, {Abul-Huda}, {Acton}, {Adams}, {Adams}, {Adler}, {Adriaensen}, {Aguilar}, {Ahmed}, {Ahmed}, {Ahmed}, {Albat}, {Albert}, {Alberts}, {Aldridge}, {Allen}, {Allen}, {Altenburg}, {Altunc}, {Alvarez}, {{\'A}lvarez-M{\'a}rquez}, {Alves de Oliveira}, {Ambrose}, {Anandakrishnan}, {Andersen}, {Anderson}, {Anderson}, {Anderson}, {Anderson}, {Aprea}, {Archer}, {Arenberg}, {Argyriou}, {Arribas}, {Artigau}, {Arvai}, {Atcheson}, {Atkinson}, {Averbukh}, {Aymergen}, {Bacinski}, {Baggett}, {Bagnasco}, {Baker}, {Balzano}, {Banks}, {Baran}, {Barker}, {Barrett}, {Barringer}, {Barto}, {Bast}, {Baudoz}, {Baum}, {Beatty}, {Beaulieu}, {Bechtold}, {Beck}, {Beddard}, {Beichman}, {Bellagama}, {Bely}, {Berger}, {Bergeron}, {Bernier}, {Bertch}, {Beskow}, {Betz}, {Biagetti}, {Birkmann}, {Bjorklund}, {Blackwood}, {Blazek}, {Blossfeld}, {Bluth}, {Boccaletti}, {Boegner}, {Bohlin}, {Boia}, {B{\"o}ker}, {Bonaventura},
  {Bond}, {Bosley}, {Boucarut}, {Bouchet}, {Bouwman}, {Bower}, {Bowers}, {Bowers}, {Boyce}, {Boyer}, {Boyer}, {Boyer}, {Boyer}, {Bradley}, {Brady}, {Brandl}, {Brannen}, {Breda}, {Bremmer}, {Brennan}, {Bresnahan}, {Bright}, {Broiles}, {Bromenschenkel}, {Brooks}, {Brooks}, {Brown}, {Brown}, {Brown}, {Bruce}, {Bryson}, {Bujanda}, {Bullock}, {Bunker}, {Bureo}, {Burt}, {Bush}, {Bushouse}, {Bussman}, {Cabaud}, {Cale}, {Calhoon}, {Calvani}, {Canipe}, {Caputo}, {Cara}, {Carey}, {Case}, {Cesari}, {Cetorelli}, {Chance}, {Chandler}, {Chaney}, {Chapman}, {Charlot}, {Chayer}, {Cheezum}, {Chen}, {Chen}, {Cherinka}, {Chichester}, {Chilton}, {Chittiraibalan}, {Clampin}, {Clark}, {Clark}, {Clark}, {Claybrooks}, {Cleveland}, {Cohen}, {Cohen}, {Col{\'o}n}, {Coleman}, {Colina}, {Comber}, {Comeau}, {Comer}, {Conde Reis}, {Connolly}, {Conroy}, {Contos}, {Contreras}, {Cook}, {Cooper}, {Cooper}, {Correia}, {Correnti}, {Cossou}, {Costanza}, {Coulais}, {Cox}, {Coyle}, {Cracraft}, {Crew}, {Curtis}, {Cusveller}, {Da Costa Maciel},
  {Dailey}, {Daugeron}, {Davidson}, {Davies}, {Davis}, {Davis}, {Day}, {de Chambure}, {de Jong}, {De Marchi}, {Dean}, {Decker}, {Delisa}, {Dell}, \& {Dellagatta}}]{Gardner2023}
{Gardner}, J.~P., {Mather}, J.~C., {Abbott}, R., {et~al.} 2023, \bibinfo{title}{{The James Webb Space Telescope Mission},} \pasp, 135, 068001, \dodoi{10.1088/1538-3873/acd1b5}

\bibitem[{V. {Gelli} {et~al.}(2025){Gelli}, {Pallottini}, {Salvadori}, {Ferrara}, {Mason}, {Carniani}, \& {Ginolfi}}]{Gelli2025}
{Gelli}, V., {Pallottini}, A., {Salvadori}, S., {et~al.} 2025, \bibinfo{title}{{Temporarily Quiescent Galaxies at Cosmic Dawn: Probing Bursty Star Formation},} \apj, 985, 126, \dodoi{10.3847/1538-4357/adc722}

\bibitem[{K. {Glazebrook} {et~al.}(1999){Glazebrook}, {Blake}, {Economou}, {Lilly}, \& {Colless}}]{Glazebrook1999}
{Glazebrook}, K., {Blake}, C., {Economou}, F., {Lilly}, S., \& {Colless}, M. 1999, \bibinfo{title}{{Measurement of the star formation rate from H{\ensuremath{\alpha}} in field galaxies at z=1},} \mnras, 306, 843, \dodoi{10.1046/j.1365-8711.1999.02576.x}

\bibitem[{N.~A. {Grogin} {et~al.}(2011){Grogin}, {Kocevski}, {Faber}, {Ferguson}, {Koekemoer}, {Riess}, {Acquaviva}, {Alexander}, {Almaini}, {Ashby}, {Barden}, {Bell}, {Bournaud}, {Brown}, {Caputi}, {Casertano}, {Cassata}, {Castellano}, {Challis}, {Chary}, {Cheung}, {Cirasuolo}, {Conselice}, {Roshan Cooray}, {Croton}, {Daddi}, {Dahlen}, {Dav{\'e}}, {de Mello}, {Dekel}, {Dickinson}, {Dolch}, {Donley}, {Dunlop}, {Dutton}, {Elbaz}, {Fazio}, {Filippenko}, {Finkelstein}, {Fontana}, {Gardner}, {Garnavich}, {Gawiser}, {Giavalisco}, {Grazian}, {Guo}, {Hathi}, {H{\"a}ussler}, {Hopkins}, {Huang}, {Huang}, {Jha}, {Kartaltepe}, {Kirshner}, {Koo}, {Lai}, {Lee}, {Li}, {Lotz}, {Lucas}, {Madau}, {McCarthy}, {McGrath}, {McIntosh}, {McLure}, {Mobasher}, {Moustakas}, {Mozena}, {Nandra}, {Newman}, {Niemi}, {Noeske}, {Papovich}, {Pentericci}, {Pope}, {Primack}, {Rajan}, {Ravindranath}, {Reddy}, {Renzini}, {Rix}, {Robaina}, {Rodney}, {Rosario}, {Rosati}, {Salimbeni}, {Scarlata}, {Siana}, {Simard}, {Smidt}, {Somerville}, {Spinrad},
  {Straughn}, {Strolger}, {Telford}, {Teplitz}, {Trump}, {van der Wel}, {Villforth}, {Wechsler}, {Weiner}, {Wiklind}, {Wild}, {Wilson}, {Wuyts}, {Yan}, \& {Yun}}]{Grogin2011}
{Grogin}, N.~A., {Kocevski}, D.~D., {Faber}, S.~M., {et~al.} 2011, \bibinfo{title}{{CANDELS: The Cosmic Assembly Near-infrared Deep Extragalactic Legacy Survey},} \apjs, 197, 35, \dodoi{10.1088/0067-0049/197/2/35}

\bibitem[{Y. {Guo} {et~al.}(2012){Guo}, {Giavalisco}, {Cassata}, {Ferguson}, {Williams}, {Dickinson}, {Koekemoer}, {Grogin}, {Chary}, {Messias}, {Tundo}, {Lin}, {Lee}, {Salimbeni}, {Fontana}, {Grazian}, {Kocevski}, {Lee}, {Villanueva}, \& {van der Wel}}]{Guo2012}
{Guo}, Y., {Giavalisco}, M., {Cassata}, P., {et~al.} 2012, \bibinfo{title}{{Rest-frame UV-Optically Selected Galaxies at 2.3 <\raisebox{-0.5ex}\textasciitilde z <\raisebox{-0.5ex}\textasciitilde 3.5: Searching for Dusty Star-forming and Passively Evolving Galaxies},} \apj, 749, 149, \dodoi{10.1088/0004-637X/749/2/149}

\bibitem[{Y. Guo {et~al.}(2016)Guo, Rafelski, Faber, Koo, Krumholz, Trump, Willner, Amorín, Barro, Bell, Gardner, Gawiser, Hathi, Koekemoer, Pacifici, Pérez-González, Ravindranath, Reddy, Teplitz, \& Yesuf}]{Guo2016}
Guo, Y., Rafelski, M., Faber, S.~M., {et~al.} 2016, \bibinfo{title}{THE BURSTY STAR FORMATION HISTORIES OF LOW-MASS GALAXIES AT 0.4 &lt; z &lt; 1 REVEALED BY STAR FORMATION RATES MEASURED FROM Hβ AND FUV,} The Astrophysical Journal, 833, 37, \dodoi{10.3847/1538-4357/833/1/37}

\bibitem[{Y. {Harikane} {et~al.}(2023){Harikane}, {Ouchi}, {Oguri}, {Ono}, {Nakajima}, {Isobe}, {Umeda}, {Mawatari}, \& {Zhang}}]{Harikane2023}
{Harikane}, Y., {Ouchi}, M., {Oguri}, M., {et~al.} 2023, \bibinfo{title}{{A Comprehensive Study of Galaxies at z 9-16 Found in the Early JWST Data: Ultraviolet Luminosity Functions and Cosmic Star Formation History at the Pre-reionization Epoch},} \apjs, 265, 5, \dodoi{10.3847/1538-4365/acaaa9}

\bibitem[{P.~F. {Hopkins} {et~al.}(2014){Hopkins}, {Kere{\v{s}}}, {O{\~n}orbe}, {Faucher-Gigu{\`e}re}, {Quataert}, {Murray}, \& {Bullock}}]{Hopkins2014}
{Hopkins}, P.~F., {Kere{\v{s}}}, D., {O{\~n}orbe}, J., {et~al.} 2014, \bibinfo{title}{{Galaxies on FIRE (Feedback In Realistic Environments): stellar feedback explains cosmologically inefficient star formation},} \mnras, 445, 581, \dodoi{10.1093/mnras/stu1738}

\bibitem[{P.~F. {Hopkins} {et~al.}(2018){Hopkins}, {Wetzel}, {Kere{\v{s}}}, {Faucher-Gigu{\`e}re}, {Quataert}, {Boylan-Kolchin}, {Murray}, {Hayward}, {Garrison-Kimmel}, {Hummels}, {Feldmann}, {Torrey}, {Ma}, {Angl{\'e}s-Alc{\'a}zar}, {Su}, {Orr}, {Schmitz}, {Escala}, {Sanderson}, {Grudi{\'c}}, {Hafen}, {Kim}, {Fitts}, {Bullock}, {Wheeler}, {Chan}, {Elbert}, \& {Narayanan}}]{Hopkins2018}
{Hopkins}, P.~F., {Wetzel}, A., {Kere{\v{s}}}, D., {et~al.} 2018, \bibinfo{title}{{FIRE-2 simulations: physics versus numerics in galaxy formation},} \mnras, 480, 800, \dodoi{10.1093/mnras/sty1690}

\bibitem[{P.~F. {Hopkins} {et~al.}(2023){Hopkins}, {Gurvich}, {Shen}, {Hafen}, {Grudi{\'c}}, {Kurinchi-Vendhan}, {Hayward}, {Jiang}, {Orr}, {Wetzel}, {Kere{\v{s}}}, {Stern}, {Faucher-Gigu{\`e}re}, {Bullock}, {Wheeler}, {El-Badry}, {Loebman}, {Moreno}, {Boylan-Kolchin}, \& {Quataert}}]{Hopkins2023}
{Hopkins}, P.~F., {Gurvich}, A.~B., {Shen}, X., {et~al.} 2023, \bibinfo{title}{{What causes the formation of discs and end of bursty star formation?},} \mnras, 525, 2241, \dodoi{10.1093/mnras/stad1902}

\bibitem[{K. {Iyer} \& E. {Gawiser}(2017){Iyer} \& {Gawiser}}]{Iyer2017}
{Iyer}, K., \& {Gawiser}, E. 2017, \bibinfo{title}{{Reconstruction of Galaxy Star Formation Histories through SED Fitting:The Dense Basis Approach},} \apj, 838, 127, \dodoi{10.3847/1538-4357/aa63f0}

\bibitem[{K.~G. {Iyer} {et~al.}(2019){Iyer}, {Gawiser}, {Faber}, {Ferguson}, {Kartaltepe}, {Koekemoer}, {Pacifici}, \& {Somerville}}]{Iyer2019}
{Iyer}, K.~G., {Gawiser}, E., {Faber}, S.~M., {et~al.} 2019, \bibinfo{title}{{Nonparametric Star Formation History Reconstruction with Gaussian Processes. I. Counting Major Episodes of Star Formation},} \apj, 879, 116, \dodoi{10.3847/1538-4357/ab2052}

\bibitem[{R.~C. {Kennicutt}(1998){Kennicutt}}]{Kennicutt1998}
{Kennicutt}, Jr., R.~C. 1998, \bibinfo{title}{{Star Formation in Galaxies Along the Hubble Sequence},} \araa, 36, 189, \dodoi{10.1146/annurev.astro.36.1.189}

\bibitem[{T. {Kimm} \& R. {Cen}(2014){Kimm} \& {Cen}}]{Kimm2014}
{Kimm}, T., \& {Cen}, R. 2014, \bibinfo{title}{{Escape Fraction of Ionizing Photons during Reionization: Effects due to Supernova Feedback and Runaway OB Stars},} \apj, 788, 121, \dodoi{10.1088/0004-637X/788/2/121}

\bibitem[{A.~M. {Koekemoer} {et~al.}(2011){Koekemoer}, {Faber}, {Ferguson}, {Grogin}, {Kocevski}, {Koo}, {Lai}, {Lotz}, {Lucas}, {McGrath}, {Ogaz}, {Rajan}, {Riess}, {Rodney}, {Strolger}, {Casertano}, {Castellano}, {Dahlen}, {Dickinson}, {Dolch}, {Fontana}, {Giavalisco}, {Grazian}, {Guo}, {Hathi}, {Huang}, {van der Wel}, {Yan}, {Acquaviva}, {Alexander}, {Almaini}, {Ashby}, {Barden}, {Bell}, {Bournaud}, {Brown}, {Caputi}, {Cassata}, {Challis}, {Chary}, {Cheung}, {Cirasuolo}, {Conselice}, {Roshan Cooray}, {Croton}, {Daddi}, {Dav{\'e}}, {de Mello}, {de Ravel}, {Dekel}, {Donley}, {Dunlop}, {Dutton}, {Elbaz}, {Fazio}, {Filippenko}, {Finkelstein}, {Frazer}, {Gardner}, {Garnavich}, {Gawiser}, {Gruetzbauch}, {Hartley}, {H{\"a}ussler}, {Herrington}, {Hopkins}, {Huang}, {Jha}, {Johnson}, {Kartaltepe}, {Khostovan}, {Kirshner}, {Lani}, {Lee}, {Li}, {Madau}, {McCarthy}, {McIntosh}, {McLure}, {McPartland}, {Mobasher}, {Moreira}, {Mortlock}, {Moustakas}, {Mozena}, {Nandra}, {Newman}, {Nielsen}, {Niemi}, {Noeske},
  {Papovich}, {Pentericci}, {Pope}, {Primack}, {Ravindranath}, {Reddy}, {Renzini}, {Rix}, {Robaina}, {Rosario}, {Rosati}, {Salimbeni}, {Scarlata}, {Siana}, {Simard}, {Smidt}, {Snyder}, {Somerville}, {Spinrad}, {Straughn}, {Telford}, {Teplitz}, {Trump}, {Vargas}, {Villforth}, {Wagner}, {Wandro}, {Wechsler}, {Weiner}, {Wiklind}, {Wild}, {Wilson}, {Wuyts}, \& {Yun}}]{Koekemoer2011}
{Koekemoer}, A.~M., {Faber}, S.~M., {Ferguson}, H.~C., {et~al.} 2011, \bibinfo{title}{{CANDELS: The Cosmic Assembly Near-infrared Deep Extragalactic Legacy Survey{\textemdash}The Hubble Space Telescope Observations, Imaging Data Products, and Mosaics},} \apjs, 197, 36, \dodoi{10.1088/0067-0049/197/2/36}

\bibitem[{V. {Kokorev} {et~al.}(2025){Kokorev}, {Ch{\'a}vez Ortiz}, {Taylor}, {Finkelstein}, {Arrabal Haro}, {Dickinson}, {Chisholm}, {Fujimoto}, {Mu{\~n}oz}, {Endsley}, {Hu}, {Napolitano}, {Wilkins}, {Akins}, {Amori{\'\i}n}, {Casey}, {Cheng}, {Cleri}, {Cole}, {Cullen}, {Daddi}, {Davis}, {Donnan}, {Dunlop}, {Fern{\'a}ndez}, {Giavalisco}, {Grogin}, {Hathi}, {Hirschmann}, {Kartaltepe}, {Koekemoer}, {Leung}, {Lucas}, {McLeod}, {Papovich}, {Pentericci}, {P{\'e}rez-Gonz{\'a}lez}, {Somerville}, {Wang}, {Yung}, \& {Zavala}}]{kokorev2025}
{Kokorev}, V., {Ch{\'a}vez Ortiz}, {\'O}.~A., {Taylor}, A.~J., {et~al.} 2025, \bibinfo{title}{{CAPERS Observations of Two UV-Bright Galaxies at z>10. More Evidence for Bursting Star Formation in the Early Universe},} arXiv e-prints, arXiv:2504.12504, \dodoi{10.48550/arXiv.2504.12504}

\bibitem[{A. {Kravtsov} \& V. {Belokurov}(2024){Kravtsov} \& {Belokurov}}]{Kravtsov2024}
{Kravtsov}, A., \& {Belokurov}, V. 2024, \bibinfo{title}{{Stochastic star formation and the abundance of $z>10$ UV-bright galaxies},} arXiv e-prints, arXiv:2405.04578, \dodoi{10.48550/arXiv.2405.04578}

\bibitem[{J. {Leja} {et~al.}(2019){Leja}, {Carnall}, {Johnson}, {Conroy}, \& {Speagle}}]{leja19}
{Leja}, J., {Carnall}, A.~C., {Johnson}, B.~D., {Conroy}, C., \& {Speagle}, J.~S. 2019, \bibinfo{title}{{How to Measure Galaxy Star Formation Histories. II. Nonparametric Models},} \apj, 876, 3, \dodoi{10.3847/1538-4357/ab133c}

\bibitem[{X. {Ma} {et~al.}(2018{\natexlab{a}}){Ma}, {Hopkins}, {Boylan-Kolchin}, {Faucher-Gigu{\`e}re}, {Quataert}, {Feldmann}, {Garrison-Kimmel}, {Hayward}, {Kere{\v{s}}}, \& {Wetzel}}]{Ma2018a}
{Ma}, X., {Hopkins}, P.~F., {Boylan-Kolchin}, M., {et~al.} 2018{\natexlab{a}}, \bibinfo{title}{{Simulating galaxies in the reionization era with FIRE-2: morphologies and sizes},} \mnras, 477, 219, \dodoi{10.1093/mnras/sty684}

\bibitem[{X. {Ma} {et~al.}(2018{\natexlab{b}}){Ma}, {Hopkins}, {Garrison-Kimmel}, {Faucher-Gigu{\`e}re}, {Quataert}, {Boylan-Kolchin}, {Hayward}, {Feldmann}, \& {Kere{\v{s}}}}]{Ma2018b}
{Ma}, X., {Hopkins}, P.~F., {Garrison-Kimmel}, S., {et~al.} 2018{\natexlab{b}}, \bibinfo{title}{{Simulating galaxies in the reionization era with FIRE-2: galaxy scaling relations, stellar mass functions, and luminosity functions},} \mnras, 478, 1694, \dodoi{10.1093/mnras/sty1024}

\bibitem[{V. {Markov} {et~al.}(2025){Markov}, {Gallerani}, {Ferrara}, {Pallottini}, {Parlanti}, {Mascia}, {Sommovigo}, \& {Kohandel}}]{Markov2025}
{Markov}, V., {Gallerani}, S., {Ferrara}, A., {et~al.} 2025, \bibinfo{title}{{The evolution of dust attenuation in z {\ensuremath{\approx}} 2-12 galaxies observed by JWST},} Nature Astronomy, 9, 458, \dodoi{10.1038/s41550-024-02426-1}

\bibitem[{C.~A. {Mason} {et~al.}(2023){Mason}, {Trenti}, \& {Treu}}]{Mason2023}
{Mason}, C.~A., {Trenti}, M., \& {Treu}, T. 2023, \bibinfo{title}{{The brightest galaxies at cosmic dawn},} \mnras, 521, 497, \dodoi{10.1093/mnras/stad035}

\bibitem[{W. {McClymont} {et~al.}(2025{\natexlab{a}}){McClymont}, {Tacchella}, {Smith}, {Kannan}, {Puchwein}, {Borrow}, {Garaldi}, {Keating}, {Vogelsberger}, {Zier}, {Shen}, {Popovic}, \& {Simmonds}}]{McClymont2025-1}
{McClymont}, W., {Tacchella}, S., {Smith}, A., {et~al.} 2025{\natexlab{a}}, \bibinfo{title}{{The THESAN-ZOOM project: Burst, quench, repeat -- unveiling the evolution of high-redshift galaxies along the star-forming main sequence},} arXiv e-prints, arXiv:2503.00106, \dodoi{10.48550/arXiv.2503.00106}

\bibitem[{W. {McClymont} {et~al.}(2025{\natexlab{b}}){McClymont}, {Tacchella}, {D'Eugenio}, {Witten}, {Ji}, {Smith}, {Maiolino}, {Arribas}, {Scholtz}, {Simmonds}, \& {Witstok}}]{McClymont2025-2}
{McClymont}, W., {Tacchella}, S., {D'Eugenio}, F., {et~al.} 2025{\natexlab{b}}, \bibinfo{title}{{The density-bounded twilight of starbursts in the early Universe},} \mnras, 540, 190, \dodoi{10.1093/mnras/staf745}

\bibitem[{V. {Mehta} {et~al.}(2023){Mehta}, {Teplitz}, {Scarlata}, {Wang}, {Alavi}, {Colbert}, {Rafelski}, {Grogin}, {Koekemoer}, {Prichard}, {Windhorst}, {Barber}, {Conselice}, {Dai}, {Gardner}, {Gawiser}, {Guo}, {Hathi}, {Arrabal Haro}, {Hayes}, {Iyer}, {Jansen}, {Ji}, {Kurczynski}, {Kuschel}, {Lucas}, {Mantha}, {O'Connell}, {Ravindranath}, {Robertson}, {Rutkowski}, {Siana}, \& {Yung}}]{Mehta2023}
{Mehta}, V., {Teplitz}, H.~I., {Scarlata}, C., {et~al.} 2023, \bibinfo{title}{{A Spatially Resolved Analysis of Star Formation Burstiness by Comparing UV and H{\ensuremath{\alpha}} in Galaxies at z {\ensuremath{\sim}} 1 with UVCANDELS},} \apj, 952, 133, \dodoi{10.3847/1538-4357/acd9cf}

\bibitem[{V. {Mehta} {et~al.}(2024){Mehta}, {Rafelski}, {Sunnquist}, {Teplitz}, {Scarlata}, {Wang}, {Fontana}, {Hathi}, {Iyer}, {Alavi}, {Colbert}, {Grogin}, {Koekemoer}, {Nedkova}, {Hayes}, {Prichard}, {Siana}, {Smith}, {Windhorst}, {Ashcraft}, {Bagley}, {Baronchelli}, {Barro}, {Blanche}, {Broussard}, {Carleton}, {Chartab}, {Codoreanu}, {Cohen}, {Conselice}, {Dai}, {Darvish}, {Dav{\'e}}, {Degroot}, {de Mello}, {Dickinson}, {Emami}, {Ferguson}, {Ferreira}, {Finkelstein}, {Finkelstein}, {Gardner}, {Gawiser}, {Gburek}, {Giavalisco}, {Grazian}, {Gronwall}, {Guo}, {Arrabal Haro}, {Hemmati}, {Howell}, {Jansen}, {Ji}, {Kaviraj}, {Kim}, {Kurczynski}, {Lazar}, {Lucas}, {MacKenty}, {Mantha}, {Martin}, {Martin}, {McCabe}, {Mobasher}, {Morales}, {O'Connell}, {Olsen}, {Otteson}, {Ravindranath}, {Redshaw}, {Rutkowski}, {Robertson}, {Sattari}, {Soto}, {Sun}, {Taamoli}, {Vanzella}, {Yung}, {Zabelle}, \& {UVCANDELS Team}}]{Mehta2024}
{Mehta}, V., {Rafelski}, M., {Sunnquist}, B., {et~al.} 2024, \bibinfo{title}{{UVCANDELS: Catalogs of Photometric Redshifts and Galaxy Physical Properties},} \apjs, 275, 17, \dodoi{10.3847/1538-4365/ad7d8f}

\bibitem[{J. {Mirocha} \& S.~R. {Furlanetto}(2023){Mirocha} \& {Furlanetto}}]{Mirocha_Furlanetto2023}
{Mirocha}, J., \& {Furlanetto}, S.~R. 2023, \bibinfo{title}{{Balancing the efficiency and stochasticity of star formation with dust extinction in z {\ensuremath{\gtrsim}} 10 galaxies observed by JWST},} \mnras, 519, 843, \dodoi{10.1093/mnras/stac3578}

\bibitem[{K.~G. Noeske {et~al.}(2007)Noeske, Weiner, Faber, Papovich, Koo, Somerville, Bundy, Conselice, Newman, Schiminovich, Le~Floc’h, Coil, Rieke, Lotz, Primack, Barmby, Cooper, Davis, Ellis, Fazio, Guhathakurta, Huang, Kassin, Martin, Phillips, Rich, Small, Willmer, \& Wilson}]{Noeske2007}
Noeske, K.~G., Weiner, B.~J., Faber, S.~M., {et~al.} 2007, \bibinfo{title}{Star Formation in AEGIS Field Galaxies since z = 1.1: The Dominance of Gradually Declining Star Formation, and the Main Sequence of Star-forming Galaxies,} The Astrophysical Journal, 660, L43, \dodoi{10.1086/517926}

\bibitem[{D.~E. {Osterbrock}(1989){Osterbrock}}]{Osterbrock1989}
{Osterbrock}, D.~E. 1989, {Astrophysics of gaseous nebulae and active galactic nuclei}

\bibitem[{D. {Pelliccia} {et~al.}(2020){Pelliccia}, {Mobasher}, {Darvish}, {Lemaux}, {Lubin}, {Hirtenstein}, {Shen}, {Wu}, {El-Badry}, {Wetzel}, \& {Jones}}]{Pelliccia2020}
{Pelliccia}, D., {Mobasher}, B., {Darvish}, B., {et~al.} 2020, \bibinfo{title}{{Effects of Stellar Feedback on Stellar and Gas Kinematics of Star-forming Galaxies at 0.6 < z < 1.0},} \apjl, 896, L26, \dodoi{10.3847/2041-8213/ab9815}

\bibitem[{ {Planck Collaboration} {et~al.}(2020){Planck Collaboration}, {Aghanim}, {Akrami}, {Ashdown}, {Aumont}, {Baccigalupi}, {Ballardini}, {Banday}, {Barreiro}, {Bartolo}, {Basak}, {Battye}, {Benabed}, {Bernard}, {Bersanelli}, {Bielewicz}, {Bock}, {Bond}, {Borrill}, {Bouchet}, {Boulanger}, {Bucher}, {Burigana}, {Butler}, {Calabrese}, {Cardoso}, {Carron}, {Challinor}, {Chiang}, {Chluba}, {Colombo}, {Combet}, {Contreras}, {Crill}, {Cuttaia}, {de Bernardis}, {de Zotti}, {Delabrouille}, {Delouis}, {Di Valentino}, {Diego}, {Dor{\'e}}, {Douspis}, {Ducout}, {Dupac}, {Dusini}, {Efstathiou}, {Elsner}, {En{\ss}lin}, {Eriksen}, {Fantaye}, {Farhang}, {Fergusson}, {Fernandez-Cobos}, {Finelli}, {Forastieri}, {Frailis}, {Fraisse}, {Franceschi}, {Frolov}, {Galeotta}, {Galli}, {Ganga}, {G{\'e}nova-Santos}, {Gerbino}, {Ghosh}, {Gonz{\'a}lez-Nuevo}, {G{\'o}rski}, {Gratton}, {Gruppuso}, {Gudmundsson}, {Hamann}, {Handley}, {Hansen}, {Herranz}, {Hildebrandt}, {Hivon}, {Huang}, {Jaffe}, {Jones}, {Karakci}, {Keih{\"a}nen},
  {Keskitalo}, {Kiiveri}, {Kim}, {Kisner}, {Knox}, {Krachmalnicoff}, {Kunz}, {Kurki-Suonio}, {Lagache}, {Lamarre}, {Lasenby}, {Lattanzi}, {Lawrence}, {Le Jeune}, {Lemos}, {Lesgourgues}, {Levrier}, {Lewis}, {Liguori}, {Lilje}, {Lilley}, {Lindholm}, {L{\'o}pez-Caniego}, {Lubin}, {Ma}, {Mac{\'\i}as-P{\'e}rez}, {Maggio}, {Maino}, {Mandolesi}, {Mangilli}, {Marcos-Caballero}, {Maris}, {Martin}, {Martinelli}, {Mart{\'\i}nez-Gonz{\'a}lez}, {Matarrese}, {Mauri}, {McEwen}, {Meinhold}, {Melchiorri}, {Mennella}, {Migliaccio}, {Millea}, {Mitra}, {Miville-Desch{\^e}nes}, {Molinari}, {Montier}, {Morgante}, {Moss}, {Natoli}, {N{\o}rgaard-Nielsen}, {Pagano}, {Paoletti}, {Partridge}, {Patanchon}, {Peiris}, {Perrotta}, {Pettorino}, {Piacentini}, {Polastri}, {Polenta}, {Puget}, {Rachen}, {Reinecke}, {Remazeilles}, {Renzi}, {Rocha}, {Rosset}, {Roudier}, {Rubi{\~n}o-Mart{\'\i}n}, {Ruiz-Granados}, {Salvati}, {Sandri}, {Savelainen}, {Scott}, {Shellard}, {Sirignano}, {Sirri}, {Spencer}, {Sunyaev}, {Suur-Uski}, {Tauber}, {Tavagnacco},
  {Tenti}, {Toffolatti}, {Tomasi}, {Trombetti}, {Valenziano}, {Valiviita}, {Van Tent}, {Vibert}, {Vielva}, {Villa}, {Vittorio}, {Wandelt}, {Wehus}, {White}, {White}, {Zacchei}, \& {Zonca}}]{Planck2020}
{Planck Collaboration}, {Aghanim}, N., {Akrami}, Y., {et~al.} 2020, \bibinfo{title}{{Planck 2018 results. VI. Cosmological parameters},} \aap, 641, A6, \dodoi{10.1051/0004-6361/201833910}

\bibitem[{P. {Popesso} {et~al.}(2023){Popesso}, {Concas}, {Cresci}, {Belli}, {Rodighiero}, {Inami}, {Dickinson}, {Ilbert}, {Pannella}, \& {Elbaz}}]{Popesso2023}
{Popesso}, P., {Concas}, A., {Cresci}, G., {et~al.} 2023, \bibinfo{title}{{The main sequence of star-forming galaxies across cosmic times},} \mnras, 519, 1526, \dodoi{10.1093/mnras/stac3214}

\bibitem[{L. {Pozzetti} {et~al.}(2010){Pozzetti}, {Bolzonella}, {Zucca}, {Zamorani}, {Lilly}, {Renzini}, {Moresco}, {Mignoli}, {Cassata}, {Tasca}, {Lamareille}, {Maier}, {Meneux}, {Halliday}, {Oesch}, {Vergani}, {Caputi}, {Kova{\v{c}}}, {Cimatti}, {Cucciati}, {Iovino}, {Peng}, {Carollo}, {Contini}, {Kneib}, {Le F{\'e}vre}, {Mainieri}, {Scodeggio}, {Bardelli}, {Bongiorno}, {Coppa}, {de la Torre}, {de Ravel}, {Franzetti}, {Garilli}, {Kampczyk}, {Knobel}, {Le Borgne}, {Le Brun}, {Pell{\`o}}, {Perez Montero}, {Ricciardelli}, {Silverman}, {Tanaka}, {Tresse}, {Abbas}, {Bottini}, {Cappi}, {Guzzo}, {Koekemoer}, {Leauthaud}, {Maccagni}, {Marinoni}, {McCracken}, {Memeo}, {Porciani}, {Scaramella}, {Scarlata}, \& {Scoville}}]{Pozzetti2010}
{Pozzetti}, L., {Bolzonella}, M., {Zucca}, E., {et~al.} 2010, \bibinfo{title}{{zCOSMOS - 10k-bright spectroscopic sample. The bimodality in the galaxy stellar mass function: exploring its evolution with redshift},} \aap, 523, A13, \dodoi{10.1051/0004-6361/200913020}

\bibitem[{N.~A. {Reddy} {et~al.}(2025){Reddy}, {Shapley}, {Sanders}, {Topping}, {Ellis}, {Pettini}, {Brammer}, {Cullen}, {Forster Schreiber}, {Khostovan}, {McLeod}, {McLure}, {Narayanan}, {Oesch}, {Pahl}, {Steidel}, \& {Berg}}]{Reddy2025}
{Reddy}, N.~A., {Shapley}, A.~E., {Sanders}, R.~L., {et~al.} 2025, \bibinfo{title}{{The JWST/AURORA Survey: Multiple Balmer and Paschen Emission Lines for Individual Star-forming Galaxies at z=1.5-4.4. I. A Diversity of Nebular Attenuation Curves and Evidence for Non-Unity Dust Covering Fractions},} arXiv e-prints, arXiv:2506.17396, \dodoi{10.48550/arXiv.2506.17396}

\bibitem[{B. {Salmon} {et~al.}(2015){Salmon}, {Papovich}, {Finkelstein}, {Tilvi}, {Finlator}, {Behroozi}, {Dahlen}, {Dav{\'e}}, {Dekel}, {Dickinson}, {Ferguson}, {Giavalisco}, {Long}, {Lu}, {Mobasher}, {Reddy}, {Somerville}, \& {Wechsler}}]{Salmon2015}
{Salmon}, B., {Papovich}, C., {Finkelstein}, S.~L., {et~al.} 2015, \bibinfo{title}{{The Relation between Star Formation Rate and Stellar Mass for Galaxies at 3.5 <= z <= 6.5 in CANDELS},} \apj, 799, 183, \dodoi{10.1088/0004-637X/799/2/183}

\bibitem[{E.~E. {Salpeter}(1955){Salpeter}}]{Salpeter1955}
{Salpeter}, E.~E. 1955, \bibinfo{title}{{The Luminosity Function and Stellar Evolution.},} \apj, 121, 161, \dodoi{10.1086/145971}

\bibitem[{S. Shen {et~al.}(2014)Shen, Madau, Conroy, Governato, \& Mayer}]{Shen2014}
Shen, S., Madau, P., Conroy, C., Governato, F., \& Mayer, L. 2014, \bibinfo{title}{THE BARYON CYCLE OF DWARF GALAXIES: DARK, BURSTY, GAS-RICH POLLUTERS,} The Astrophysical Journal, 792, 99, \dodoi{10.1088/0004-637X/792/2/99}

\bibitem[{X. {Shen} {et~al.}(2023){Shen}, {Vogelsberger}, {Boylan-Kolchin}, {Tacchella}, \& {Kannan}}]{Shen2023}
{Shen}, X., {Vogelsberger}, M., {Boylan-Kolchin}, M., {Tacchella}, S., \& {Kannan}, R. 2023, \bibinfo{title}{{The impact of UV variability on the abundance of bright galaxies at z {\ensuremath{\geq}} 9},} \mnras, 525, 3254, \dodoi{10.1093/mnras/stad2508}

\bibitem[{I. {Shivaei} {et~al.}(2025){Shivaei}, {Naidu}, {Rodr{\'\i}guez Montero}, {Matsumoto}, {Leja}, {Matthee}, {Johnson}, {Oesch}, {Chevallard}, {Adamo}, {Bodansky}, {Bunker}, {Covelo Paz}, {Di Cesare}, {Egami}, {Furtak}, {Heintz}, {Kramarenko}, {Meyer}, {Reddy}, {Rinaldi}, {Tacchella}, {Torralba}, {Witstok}, {Wozniak}, \& {Xiao}}]{Shivaei2025}
{Shivaei}, I., {Naidu}, R.~P., {Rodr{\'\i}guez Montero}, F., {et~al.} 2025, \bibinfo{title}{{The Diversity and Evolution of Dust Attenuation Curves from Redshift z \raisebox{-0.5ex}\textasciitilde 1 to 9},} arXiv e-prints, arXiv:2509.01795, \dodoi{10.48550/arXiv.2509.01795}

\bibitem[{R.~S. {Somerville} {et~al.}(2025){Somerville}, {Yung}, {Lancaster}, {Menon}, {Sommovigo}, \& {Finkelstein}}]{Somerville2025}
{Somerville}, R.~S., {Yung}, L.~Y.~A., {Lancaster}, L., {et~al.} 2025, \bibinfo{title}{{Density modulated star formation efficiency: implications for the observed abundance of ultra-violet luminous galaxies at z>10},} arXiv e-prints, arXiv:2505.05442, \dodoi{10.48550/arXiv.2505.05442}

\bibitem[{M. {Sparre} {et~al.}(2017){Sparre}, {Hayward}, {Feldmann}, {Faucher-Gigu{\`e}re}, {Muratov}, {Kere{\v{s}}}, \& {Hopkins}}]{Sparre2017}
{Sparre}, M., {Hayward}, C.~C., {Feldmann}, R., {et~al.} 2017, \bibinfo{title}{{(Star)bursts of FIRE: observational signatures of bursty star formation in galaxies},} \mnras, 466, 88, \dodoi{10.1093/mnras/stw3011}

\bibitem[{J.~S. {Speagle} {et~al.}(2014){Speagle}, {Steinhardt}, {Capak}, \& {Silverman}}]{Speagle2014}
{Speagle}, J.~S., {Steinhardt}, C.~L., {Capak}, P.~L., \& {Silverman}, J.~D. 2014, \bibinfo{title}{{A Highly Consistent Framework for the Evolution of the Star-Forming ``Main Sequence'' from z \raisebox{-0.5ex}\textasciitilde 0-6},} \apjs, 214, 15, \dodoi{10.1088/0067-0049/214/2/15}

\bibitem[{M. {Sullivan} {et~al.}(2001){Sullivan}, {Mobasher}, {Chan}, {Cram}, {Ellis}, {Treyer}, \& {Hopkins}}]{Sullivan2001}
{Sullivan}, M., {Mobasher}, B., {Chan}, B., {et~al.} 2001, \bibinfo{title}{{A Comparison of Independent Star Formation Diagnostics for an Ultraviolet-selected Sample of Nearby Galaxies},} \apj, 558, 72, \dodoi{10.1086/322451}

\bibitem[{M. {Sullivan} {et~al.}(2000){Sullivan}, {Treyer}, {Ellis}, {Bridges}, {Milliard}, \& {Donas}}]{Sullivan2000}
{Sullivan}, M., {Treyer}, M.~A., {Ellis}, R.~S., {et~al.} 2000, \bibinfo{title}{{An ultraviolet-selected galaxy redshift survey - II. The physical nature of star formation in an enlarged sample},} \mnras, 312, 442, \dodoi{10.1046/j.1365-8711.2000.03140.x}

\bibitem[{G. {Sun} {et~al.}(2023a){Sun}, {Faucher-Gigu{\`e}re}, {Hayward}, \& {Shen}}]{Sun2023a}
{Sun}, G., {Faucher-Gigu{\`e}re}, C.-A., {Hayward}, C.~C., \& {Shen}, X. 2023a, \bibinfo{title}{{Seen and unseen: bursty star formation and its implications for observations of high-redshift galaxies with JWST},} \mnras, 526, 2665, \dodoi{10.1093/mnras/stad2902}

\bibitem[{G. Sun {et~al.}(2023b)Sun, Faucher-Giguère, Hayward, Shen, Wetzel, \& Cochrane}]{Sun2023b}
Sun, G., Faucher-Giguère, C.-A., Hayward, C.~C., {et~al.} 2023b, \bibinfo{title}{Bursty Star Formation Naturally Explains the Abundance of Bright Galaxies at Cosmic Dawn,} The Astrophysical Journal Letters, 955, L35, \dodoi{10.3847/2041-8213/acf85a}

\bibitem[{S. {Tacchella} {et~al.}(2016){Tacchella}, {Dekel}, {Carollo}, {Ceverino}, {DeGraf}, {Lapiner}, {Mandelker}, \& {Primack Joel}}]{Tacchella2016}
{Tacchella}, S., {Dekel}, A., {Carollo}, C.~M., {et~al.} 2016, \bibinfo{title}{{The confinement of star-forming galaxies into a main sequence through episodes of gas compaction, depletion and replenishment},} \mnras, 457, 2790, \dodoi{10.1093/mnras/stw131}

\bibitem[{S. {Tacchella} {et~al.}(2020){Tacchella}, {Forbes}, \& {Caplar}}]{Tacchella2020}
{Tacchella}, S., {Forbes}, J.~C., \& {Caplar}, N. 2020, \bibinfo{title}{{Stochastic modelling of star-formation histories II: star-formation variability from molecular clouds and gas inflow},} \mnras, 497, 698, \dodoi{10.1093/mnras/staa1838}

\bibitem[{S. {Tacchella} {et~al.}(2022{\natexlab{a}}){Tacchella}, {Smith}, {Kannan}, {Marinacci}, {Hernquist}, {Vogelsberger}, {Torrey}, {Sales}, \& {Li}}]{Tacchella2022-2}
{Tacchella}, S., {Smith}, A., {Kannan}, R., {et~al.} 2022{\natexlab{a}}, \bibinfo{title}{{H {\ensuremath{\alpha}} emission in local galaxies: star formation, time variability, and the diffuse ionized gas},} \mnras, 513, 2904, \dodoi{10.1093/mnras/stac818}

\bibitem[{S. {Tacchella} {et~al.}(2022{\natexlab{b}}){Tacchella}, {Finkelstein}, {Bagley}, {Dickinson}, {Ferguson}, {Giavalisco}, {Graziani}, {Grogin}, {Hathi}, {Hutchison}, {Jung}, {Koekemoer}, {Larson}, {Papovich}, {Pirzkal}, {Rojas-Ruiz}, {Song}, {Schneider}, {Somerville}, {Wilkins}, \& {Yung}}]{Tacchella2022-1}
{Tacchella}, S., {Finkelstein}, S.~L., {Bagley}, M., {et~al.} 2022{\natexlab{b}}, \bibinfo{title}{{On the Stellar Populations of Galaxies at z = 9-11: The Growth of Metals and Stellar Mass at Early Times},} \apj, 927, 170, \dodoi{10.3847/1538-4357/ac4cad}

\bibitem[{S. {Tacchella} {et~al.}(2023){Tacchella}, {Eisenstein}, {Hainline}, {Johnson}, {Baker}, {Helton}, {Robertson}, {Suess}, {Chen}, {Nelson}, {Pusk{\'a}s}, {Sun}, {Alberts}, {Egami}, {Hausen}, {Rieke}, {Rieke}, {Shivaei}, {Williams}, {Willmer}, {Bunker}, {Cameron}, {Carniani}, {Charlot}, {Curti}, {Curtis-Lake}, {Looser}, {Maiolino}, {Maseda}, {Rawle}, {Rix}, {Smit}, {{\"U}bler}, {Willott}, {Witstok}, {Baum}, {Bhatawdekar}, {Boyett}, {Danhaive}, {de Graaff}, {Endsley}, {Ji}, {Lyu}, {Sandles}, {Saxena}, {Scholtz}, {Topping}, \& {Whitler}}]{tacchella23}
{Tacchella}, S., {Eisenstein}, D.~J., {Hainline}, K., {et~al.} 2023, \bibinfo{title}{{JADES Imaging of GN-z11: Revealing the Morphology and Environment of a Luminous Galaxy 430 Myr after the Big Bang},} \apj, 952, 74, \dodoi{10.3847/1538-4357/acdbc6}

\bibitem[{A.~J. {Taylor} {et~al.}(2025){Taylor}, {Finkelstein}, {Kocevski}, {Jeon}, {Bromm}, {Amor{\'\i}n}, {Arrabal Haro}, {Backhaus}, {Bagley}, {Banados}, {Bhatawdekar}, {Brooks}, {Calabr{\`o}}, {Ch{\'a}vez Ortiz}, {Cheng}, {Cleri}, {Cole}, {Davis}, {Dickinson}, {Donnan}, {Dunlop}, {Ellis}, {Fern{\'a}ndez}, {Fontana}, {Fujimoto}, {Giavalisco}, {Grazian}, {Guo}, {Hathi}, {Holwerda}, {Hirschmann}, {Inayoshi}, {Kartaltepe}, {Khusanova}, {Koekemoer}, {Kokorev}, {Larson}, {Leung}, {Lucas}, {McLeod}, {Napolitano}, {Onoue}, {Pacucci}, {Papovich}, {P{\'e}rez-Gonz{\'a}lez}, {Pirzkal}, {Somerville}, {Trump}, {Wilkins}, {Yung}, \& {Zhang}}]{Taylor2025}
{Taylor}, A.~J., {Finkelstein}, S.~L., {Kocevski}, D.~D., {et~al.} 2025, \bibinfo{title}{{Broad-line AGNs at 3.5 < z < 6: The Black Hole Mass Function and a Connection with Little Red Dots},} \apj, 986, 165, \dodoi{10.3847/1538-4357/add15b}

\bibitem[{P.~A.~M. {van Hoof}(2018){van Hoof}}]{Hoof2018}
{van Hoof}, P. A.~M. 2018, \bibinfo{title}{{Recent Development of the Atomic Line List},} Galaxies, 6, 63, \dodoi{10.3390/galaxies6020063}

\bibitem[{P. {Virtanen} {et~al.}(2020){Virtanen}, {Gommers}, {Oliphant}, {Haberland}, {Reddy}, {Cournapeau}, {Burovski}, {Peterson}, {Weckesser}, {Bright}, {van der Walt}, {Brett}, {Wilson}, {Millman}, {Mayorov}, {Nelson}, {Jones}, {Kern}, {Larson}, {Carey}, {Polat}, {Feng}, {Moore}, {VanderPlas}, {Laxalde}, {Perktold}, {Cimrman}, {Henriksen}, {Quintero}, {Harris}, {Archibald}, {Ribeiro}, {Pedregosa}, {van Mulbregt}, \& {SciPy 1. 0 Contributors}}]{Virtanen2020}
{Virtanen}, P., {Gommers}, R., {Oliphant}, T.~E., {et~al.} 2020, \bibinfo{title}{{SciPy 1.0: fundamental algorithms for scientific computing in Python},} Nature Methods, 17, 261, \dodoi{10.1038/s41592-019-0686-2}

\bibitem[{B. {Wang} {et~al.}(2024){Wang}, {Leja}, {de Graaff}, {Brammer}, {Weibel}, {van Dokkum}, {Baggen}, {Suess}, {Greene}, {Bezanson}, {Cleri}, {Hirschmann}, {Labb{\'e}}, {Matthee}, {McConachie}, {Naidu}, {Nelson}, {Oesch}, {Setton}, \& {Williams}}]{Wang2024}
{Wang}, B., {Leja}, J., {de Graaff}, A., {et~al.} 2024, \bibinfo{title}{{RUBIES: Evolved Stellar Populations with Extended Formation Histories at z {\ensuremath{\sim}} 7{\textendash}8 in Candidate Massive Galaxies Identified with JWST/NIRSpec},} \apjl, 969, L13, \dodoi{10.3847/2041-8213/ad55f7}

\bibitem[{E.~J. {Weller} {et~al.}(2025){Weller}, {Pacucci}, {Ni}, {Hernquist}, \& {Park}}]{Weller2025}
{Weller}, E.~J., {Pacucci}, F., {Ni}, Y., {Hernquist}, L., \& {Park}, M. 2025, \bibinfo{title}{{Discrepancies between JWST Observations and Simulations of Quenched Massive Galaxies at z > 3: A Comparative Study with IllustrisTNG and ASTRID},} \apj, 979, 181, \dodoi{10.3847/1538-4357/ada360}

\bibitem[{K.~E. {Whitaker} {et~al.}(2014){Whitaker}, {Franx}, {Leja}, {van Dokkum}, {Henry}, {Skelton}, {Fumagalli}, {Momcheva}, {Brammer}, {Labb{\'e}}, {Nelson}, \& {Rigby}}]{Whitaker2014}
{Whitaker}, K.~E., {Franx}, M., {Leja}, J., {et~al.} 2014, \bibinfo{title}{{Constraining the Low-mass Slope of the Star Formation Sequence at 0.5 < z < 2.5},} \apj, 795, 104, \dodoi{10.1088/0004-637X/795/2/104}

\bibitem[{L. {Whitler} {et~al.}(2025){Whitler}, {Stark}, {Topping}, {Robertson}, {Rieke}, {Hainline}, {Endsley}, {Chen}, {Baker}, {Bhatawdekar}, {Bunker}, {Carniani}, {Charlot}, {Chevallard}, {Curtis-Lake}, {Egami}, {Eisenstein}, {Helton}, {Ji}, {Johnson}, {P{\'e}rez-Gonz{\'a}lez}, {Rinaldi}, {Tacchella}, {Williams}, {Willmer}, {Willott}, \& {Witstok}}]{Whitler2025}
{Whitler}, L., {Stark}, D.~P., {Topping}, M.~W., {et~al.} 2025, \bibinfo{title}{{The $z rsim 9$ galaxy UV luminosity function from the JWST Advanced Deep Extragalactic Survey: insights into early galaxy evolution and reionization},} arXiv e-prints, arXiv:2501.00984, \dodoi{10.48550/arXiv.2501.00984}

\bibitem[{S.~M. {Wilkins} {et~al.}(2023){Wilkins}, {Vijayan}, {Lovell}, {Roper}, {Zackrisson}, {Irodotou}, {Seeyave}, {Kuusisto}, {Thomas}, {Caruana}, \& {Conselice}}]{Wilkins2023}
{Wilkins}, S.~M., {Vijayan}, A.~P., {Lovell}, C.~C., {et~al.} 2023, \bibinfo{title}{{First Light And Reionization Epoch Simulations (FLARES) VII: The star formation and metal enrichment histories of galaxies in the early Universe},} \mnras, 518, 3935, \dodoi{10.1093/mnras/stac3281}

\end{thebibliography}
\bibliographystyle{aasjournalv7}

\end{document}